\documentclass[a4paper,fleqn,usenatbib]{mnras}
\usepackage[T1]{fontenc}
\usepackage{ae,aecompl}
\usepackage{amsmath}
\usepackage{amsfonts}
\usepackage{amssymb}
\usepackage{graphicx}
\usepackage{epstopdf}
\usepackage[pdf]{pstricks}
\DeclareGraphicsExtensions{.eps}
\def\mnras{MNRAS}
\def\aap{A$\&$A}

\def\apj{ApJ}
\def\apjl{ApJ Letters}
\def\apjs{ApJS}

\def\icarus{Icarus}
\def\planss{Planetary and Space Science}
\newcommand{\mmsn}{{\rm MMSN}}
\newcommand{\me}{\, {\rm M}_{\oplus}}
\newcommand{\Rearth}{\, {\rm R}_{\oplus}}
\newcommand{\au}{\, {\rm au}}
\title[In situ accretion of gaseous envelopes]{In situ accretion of gaseous envelopes on to planetary cores embedded in evolving protoplanetary discs}
\author[G. A. L. Coleman et al]{Gavin A. L. Coleman$^{1}$\thanks{Email: g.coleman@qmul.ac.uk},
John C. B. Papaloizou$^{2}$,
Richard P. Nelson$^{1,3}$\\
$^{1}$Astronomy Unit, Queen Mary University of London, Mile End Road, London, E1 4NS, U.K.\\
$^{2}$DAMTP, University of Cambridge, Wilberforce Road, Cambridge, CB3 0WA, U.K.\\
$^{3}$Kavli Institute for Theoretical Physics, University of California, Santa Barbara, CA 93106, USA.}

\date{}
\begin{document}
\maketitle
\begin{abstract}
The core accretion hypothesis posits that planets with significant gaseous envelopes accreted them from their protoplanetary discs after the formation of rocky/icy cores.
Observations indicate that such exoplanets exist at a broad range of orbital radii,
but it is not known whether they accreted their envelopes in situ, or originated elsewhere and migrated to their current locations.
We consider the evolution of solid cores embedded in evolving viscous discs that undergo gaseous envelope accretion in situ with orbital radii in the range 0.1-$10\au$.
Additionally, we determine the long-term evolution of the planets that had no runaway gas accretion phase after disc dispersal.
We find: (i) Planets with $5 \me$ cores never undergo runaway accretion.
The most massive envelope contained $2.8 \me$ with the planet orbiting at $10 \au$.
(ii) Accretion is more efficient onto $10 \me$ and  $15 \me$ cores.
For orbital radii $a_{\rm p} \ge 0.5 \au$, $15 \me$ cores always experienced runaway gas accretion.
For $a_{\rm p} \ge 5 \au$, all but one of the $10 \me$ cores experienced runaway gas accretion.
No planets experienced runaway growth at $a_{\rm p} = 0.1 \au$.
(iii) We find that, after disc dispersal, planets with significant gaseous envelopes cool and contract on Gyr time-scales, the contraction time being sensitive to the opacity assumed.
Our results indicate that Hot Jupiters with core masses $\lesssim 15 \me$ at $\lesssim 0.1 \au$ likely accreted their gaseous envelopes at larger distances and migrated inwards.
Consistently with the known exoplanet population, Super-Earths and mini-Neptunes at small radii during the disc lifetime, accrete only modest gaseous envelopes.
\end{abstract}

\section{Introduction}
\label{sec:intro}
Two distinct paradigms have been proposed to explain the origin of gas giant planets: core nucleated instability \citep{Perri1974, Mizuno1978, Stevenson1982} and gravitational instability \citep[e.g.][and references therein]{Helled14}. In principle, given the presence of a solid core of sufficient mass, core nucleated instability can lead to the formation of gas giant planets at almost any orbital location within a protoplanetary disc during its lifetime.

For gravitational instability to operate it is required that the disc gas can cool efficiently on orbital time-scales.
Accordingly this mode of planet formation is most likely to occur at large stellocentric distances ($\gtrsim 50 \au$) where cooling times are short \citep[e.g.][]{Gammie2001}.
On the other hand, formation of more massive objects such as brown dwarfs may be preferred \citep[see e.g.][]{Kratter2016}.

In this paper, we examine the accretion of gas on to already formed cores located at a wide variety of stellocentric distances, ranging between $0.1\au$ and $10 \au$ within evolving protoplanetary disc models, to determine where gas giant planets can form via in situ gas accretion, and the likely envelope masses of planets that do not achieve runaway growth.
Most previous studies of gas accretion on to planetary cores have used either static models of planetary envelopes, where there is a thermal equilibrium for which planetesimal accretion on to the core provides the energy source that balances radiative losses from the envelope \citep{Perri1974, Mizuno1978, Rafikov2006}, or have used quasi-static models where accretion and Kelvin-Helmholtz contraction of a gaseous envelope proceeds through a sequence of phases for which the model is assumed to be in hydrostatic equilibrium \citep{BodenheimerPollack86,Pollack}. In the former case, for a given planetesimal accretion luminosity, there is a core mass above which an envelope in hydrostatic and thermal equilibrium cannot exist, and it is normally assumed that once this critical core mass has been exceeded then rapid gas accretion will ensue. This provides one definition of the critical core mass that is often cited in the literature. However in the latter case, the accretion and settling of gas provides a source of gravitational energy that replaces the energy that is radiated away by the planet when there is no planetesimal accretion. Continuing radiative losses then enable further settling and gas accretion to occur. The quasi-static models show that giant planet formation goes through a number of phases that are characterised by different time-scales. The core is normally built by rapid planetesimal accretion in phase 1. In phase 2, the planetesimal accretion rate drops and the atmosphere grows slowly, where the rate of gas accretion is controlled by the rate of the envelope's Kelvin-Helmholtz contraction. As the envelope mass approaches that contained in the core (the so-called cross-over mass), phase 3 begins and gas accretion increases quasi-exponentially because of the increasing importance of the envelope self-gravity, leading to rapid growth of the planet.

In this paper we adopt the approach developed in \citet{PapNelson2005} to examine the accretion of gaseous envelopes onto planet cores using a sequence of quasi-static models. When discussing the critical core mass in this work, we refer to the fact that the sequence of hydrostatic models that we compute eventually comes to an end when a hydrostatic model cannot be obtained for a given core and envelope mass that has a luminosity that is sourced by the contraction and continuing accretion of the envelope. At this stage the core is said to be of a critical mass. We assume that at this point the planet enters runaway gas accretion and will eventually detach from the disc as the envelope contracts to become smaller than the planet Hill sphere.
We also consider the long term evolution of the models that do not undergo runaway gas accretion after the gas disc has dispersed by evolving them for up to Gyr time-scales in the absence of gas accretion and under the assumption that the planet now has a free surface through which it radiates rather than being embedded in the protoplanetary disc.

Our motivation for this work is manifold. First, we have incorporated the gas accretion models of \cite{PapNelson2005} into the combined N-body and protoplanetary disc evolution code presented in \cite{ColemanNelson14, ColemanNelson16}, in order to increase the realism of the planetary system formation simulations that can be computed with this numerical tool. The purpose of this paper is to provide a simplified examination of gas accretion onto isolated, non-interacting planetary cores at a variety of orbital locations using the evolving disc models from \cite{ColemanNelson16}, as a precursor to presenting simulations that consider the contemporaneous accretion of solids and gas onto protoplanets under the influence of migration within the evolving disc models. A second motivation is to examine how accreting planets with different initial core masses at different locations in the disc evolve within disc lifetimes. Observations indicate the presence of exoplanets (super-Earths, Neptunes and Jupiters) that are likely to be gas bearing with a broad range of orbital radii, covering the interval $0.04 \au \lesssim a_{\rm p} \lesssim 100 \au$ \citep[e.g.][]{Borucki2011, Marois2008}. Analysis of the occurrence rates as a function of planet radius in the \emph{Kepler} data suggests that planets with radii in the range $0.85 \le R_{\rm p} \le 4 \Rearth$ are present around $\sim 50\%$ of Sun-like stars \citep[e.g.][]{Fressin2013}, which is consonant with the results from earlier radial velocity surveys that focussed on low mass planets \citep{Mayor2011}. Transit timing variations \citep{WuLithwick2013,HaddenLithwick2016} and radial velocity measurements \citep{Marcy2014,Weiss2016} indicate that the super-Earths and Neptunes in the \emph{Kepler} data with radii $R_{\rm p} < 4 \Rearth$ have masses in the range $2 \lesssim M_{\rm p} \lesssim 15 \me$, and show a tendency for the bulk density to increase for $R_{\rm p} < 1.5 \Rearth$ and decrease for $1.5 < R_{\rm p} < 4 \Rearth$, indicating that the smaller planets are composed almost entirely of solids whereas the larger planets have larger radii due to the presence of significant gaseous envelopes that have been accreted during formation. Comparing theoretical models with the data suggests that the gaseous envelopes contribute little to the total masses of the super-Earths and Neptunes \citep{LopezFortney2014}, with recent estimates indicating that planets with $R_{\rm p} \le 1.2 \Rearth$ contain $\sim 1\%$ gas by mass, rising to $\sim 5\%$ for planets with $R_{\rm p} \sim 3 \Rearth$ \citep{WolfgangLopez2015}. Considering the gas giant exoplanets, numerous attempts have been made to constrain the core masses and total heavy element abundances of those transiting planets that have mass estimates from radial velocity measurements using evolutionary models of the internal structure \citep{Guillot2006,Burrows2007,MillerFortney2011,Thorngren2016}. These studies provide a strong indication that the majority of gas giant exoplanets are substantially enriched with heavy elements relative to their host stars, with heavy element masses typically being $\sim 20 \me$ or larger. These results are consistent with the idea that giant planets form when a sufficiently massive (i.e. in the range 10 -- $15 \me$) core forms that is able to accrete gas and undergo runaway gas accretion sufficiently early in the disc lifetime, in basic agreement with the theoretical core accretion models. In the absence of a sufficiently massive core that forms early enough, a more common outcome is the formation of super-Earths or Neptunes that fail to become gas giants because of the relatively slow rate of gas accretion until the envelope mass becomes comparable to the core mass. It is clearly of interest to address the above observational results through theoretical modelling of in situ planet formation, and to compare the outcomes of our calculations with recent work that has considered gas accretion onto planets with short \citep{Lee2014,Batygin2016} and long orbital periods \citep{Piso2014,Piso2015}. Our third motivation is to consider the longer term cooling and contraction of the gaseous envelopes for those planets that remain in a state of quasi-static equilibrium throughout the period of gas accretion when the disc is present. Since the observations are typically relevant to exoplanets that have existed around their stars for Gyrs, they will have evolved for time periods of this duration after the epoch of formation. This is the first time that our models have been used to examine this longer term evolution, and our approach is similar to that used in recent population synthesis studies \citep[see the recent review by][]{Mordasini2015}.

This paper is organised as follows. We present the basic equations and their methods of solution in Sections~\ref{sec:atmos} and \ref{sec:freecalc}. In Section~\ref{sec:discs} we present the evolution of the disc models, and in Section~\ref{sec:insitu} we present calculations of the accretion of gas onto cores of varying mass located at different disc locations. In Section~\ref{sec:lte} we present calculations of the long term evolution of planetary envelopes, and in Section~\ref{sec:conc} we discuss our results and draw conclusions.

\section{Calculating gas accretion rates}
\label{sec:atmos}
The temporal evolution of the accreting gaseous envelopes in our models are computed using the methods described in \cite{PapNelson2005}. This adopts the approach presented in previous works \citep[e.g.][]{Pap-Terquem-envelopes,Pollack,BodenheimerPollack86} in approximating the gaseous envelope structure as being non-rotating, spherically symmetric and in hydrostatic equilibrium, similar to the approach used in modelling stellar evolution. The equation of hydrostatic balance is given by
\begin{equation}
\label{eq:hydrostatic}
\frac{dP}{dR}=-\rho g
\end{equation}
where $P$ is the pressure, $\rho$ is the density, and $g$ is the acceleration due to gravity, $g = GM(R)/R^2$ with $M(R)$ being the mass, including that of the solid core, interior to the radius $R$, and $G$ is is the gravitational constant.

We adopt the equation of state for a hydrogen and helium mixture given by \citet{Saumon95}. The mass fractions of hydrogen and helium are 0.7 and 0.28 respectively. The luminosity, $L_{\rm rad}$, transported by radiation is given by
\begin{equation}
L_{\rm rad} = -\frac{64\pi\sigma R^2 T^3}{3\kappa\rho}\dfrac{dT}{dR}
\end{equation}
where $\sigma$ is the Stefan-Boltzmann constant, $T$ is the temperature, and $\kappa$ is the opacity.
We assume that the opacity is given by the Rosseland mean opacity, with the temperature and density dependencies being calculated using the formulae in \citet{Bell97} for temperatures below 3730 K, and by \citet{Bell94} above 3730 K.

\subsection{Inner convection zones}
When the planet is radiating on account of the envelope settling onto the core, most of the mass in the envelope is unstable to convection so a theory of energy transport by convection is required. We adopt the conventional mixing length theory \citep[e.g.][]{Cox68}.

The radiative and adiabatic temperature gradients $\nabla_{\rm rad}$ and $\nabla_{\rm ad}$ are defined as
\begin{equation}
\nabla_{\rm rad} = \frac{3\kappa L_{\rm tot}P}{64\pi\sigma GMT^4}
\end{equation}
and
\begin{equation}
\nabla_{\rm ad} = \left(\dfrac{\partial {\rm ln} T}{\partial {\rm ln} P}\right)_{\rm S},
\end{equation}
with the subscript S denoting evaluation at constant entropy.
The total luminosity is $L_{\rm tot}$. Within our models this is equal to the rate of release of gravitational energy of gas accreted from the surrounding protoplanetary disc, as well as that liberated from the gravitational settling of the gaseous envelope (see sect. \ref{sec:atmos_accretion}).
We include the binding energy of the planet core in the initial calculation of the luminosity, and in future work we will include the accretion luminosity obtained from the release of gravitational energy of accreted planetesimals. In this paper, however, the accretion of solids in the form of planetesimals, boulders or pebbles is neglected. 

When the gas is convectively stable, such that $\nabla_{\rm rad} < \nabla_{\rm ad}$, the energy is transported entirely through radiation. 
However when $\nabla_{\rm rad} > \nabla_{\rm ad}$, the gas is convectively unstable and so some of the energy is transported through convection.
The total luminosity passing through a sphere of radius $r$ is equal to $L_{\rm tot}(r) = L_{\rm rad} +  L_{\rm conv}$, where $L_{\rm conv}$ is the luminosity associated with convection.
Adopting the mixing length theory \citep{Cox68}, this is equal to
\begin{equation}
\label{eq:lconv}
L_{\rm conv} = \pi R^2C_p\Lambda^2\left[\left(\dfrac{dT}{dR}\right)_{\rm S}-\left(\dfrac{dT}{dR}\right)\right]^{3/2}\times\sqrt{\frac{1}{2}\rho g \left|\left(\dfrac{\partial \rho}{\partial T}\right)_{\rm P}\right|}
\end{equation}
where $\Lambda = \left|\alpha P/(dP/dR)\right|$ is the mixing length, $\alpha$ being a constant parameter expected to be of order unity, $(dT/dR)_{\rm S}=\nabla_{\rm ad}T(d {\rm ln}P/dR)$, and the subscript P denotes evaluation at constant pressure.
We adopt the mixing length parameter $\alpha = 1$.
As for our models, convection is efficient near the core where the gravitational energy is liberated, the temperature gradient will be close to the adiabatic gradient with the luminosity dropping out of the equations.
For this reason we take $L_{\rm tot}$ to be constant which is the approximation usually made when considering stellar envelopes in corresponding circumstances.

\subsection{Boundary conditions}
\label{sec:boundary}
\subsubsection{Inner boundary}
We assume that there is a solid core of mass $M_{\rm core}$ with a uniform density, $\rho_{\rm core} = 3.2$ gcm$^{-3}$ \citep{BodenheimerPollack86}.
The models that calculate the structure of the gaseous envelope must therefore satisfy that the total mass $M(R_{\rm core})=M_{\rm core}$ when
\begin{equation}
R = R_{\rm core} = \left(\frac{3M_{\rm core}}{4\pi\rho_{\rm core}}\right)^{1/3}.
\end{equation}

\subsubsection{Outer boundary}
We assume that the envelope structure extends to the Roche lobe, or the boundary of the Hill sphere beyond which material must be gravitationally unbound from the planet. The surface radius of the core-envelope structure, $R_{\rm surf}$, is then given by 
\begin{equation}
R_{\rm surf}=\frac{2}{3}\left(\frac{M_{\rm p}}{3M_*}\right)^{1/3}a_{\rm p}
\end{equation}
where $M_{\rm p}$ is the total planet mass (gaseous envelope and solid core), and $a_{\rm p}$ is the orbital radius of the planet.
We assume that the outer layers of the gaseous envelope eventually join smoothly to the surrounding protoplanetary disc, where $T = T_{\rm disc}$, $P = P_{\rm disc}$ and $\rho = \rho_{\rm disc}$ respectively.
Therefore the boundary conditions are that at $R = R_{\rm surf}$, $M(R_{\rm surf}) = M_{\rm p}$, $P_{\rm surf} = P_{\rm disc}$ and the temperature is given by
\begin{equation}
T_{\rm surf} = (T_{\rm disc}^4+T^4_{\rm effp})^{1/4}
\label{eq:Tsurf}
\end{equation}
where $T^4_{\rm effp} = 3\tau_{\rm surf}L_{\rm tot}/(16\pi\sigma R^2_{\rm surf})$ with $L_{\rm tot}$ denoting the total luminosity escaping from the surface.
We approximate the additional optical depth above the planet's atmosphere, through which radiation passes, by \citep{Pap-Terquem-envelopes}
\begin{equation}
\tau_{\rm surf}=\kappa(\rho_{\rm disc}T_{\rm disc})\rho_{\rm disc}R_{\rm surf}.
\end{equation}
Equation~\ref{eq:Tsurf} expresses the fact that $T_{\rm surf}$ must exceed the disc temperature $T_{\rm disc}$ at $R = R_{\rm surf}$ in order that the luminosity be radiated away from the planet and into the surrounding protoplanetary disc.

\subsection{Accretion, settling and envelope evolution sequences}
\label{sec:atmos_accretion}
The models we present here provide an evolutionary sequence where the mass $M_{\rm p}$ increases through the accretion of gas from the surrounding protoplanetary disc.
As the gas settles, it also liberates gravitational energy, increasing the planet's luminosity.
To calculate this energy attributed to the planet's evolution, we consider the total energy of the planet within the Roche lobe
\begin{equation}
E = \int^{M_{\rm p}}_0 \left(U-\frac{GM}{R}\right)dM.
\label{eq:envenergy}
\end{equation}
Here $U$ is the internal energy per unit mass and we neglect the energy involved in bringing material from infinity to the Roche lobe.
We justify this as the majority of the mass is concentrated well inside the Roche lobe, close to the planet's core where the specific energies are significantly higher.

Since the planet radiates energy into the surrounding disc, this energy must come from either the settling of the atmosphere described above, or through the release of gravitational energy from gas that is accreted by the planet.
This change in the energy of the planet as its mass increases is $dE = (dE/dM_{\rm p})dM_{\rm p}$.
As this energy change balances the loss of energy through radiation, then the conservation of energy requires that
\begin{equation}
\label{eq:dmpdtequation}
\dfrac{dE}{dM_{\rm p}}\dfrac{dM_{\rm p}}{dt}=-L,
\end{equation}
with $L$ being the luminosity at the surface.
This then determines the evolution of the atmosphere as it settles and continues to accrete gas from the surrounding disc.
We note that the accretion of planetesimals, neglected in this work, can also provide an additional energy source. If included in the models, then this would reduce the gas accretion rate as it would supply some fraction of the energy required to balance that radiated by the planet. The final masses of the gas envelopes would then be reduced accordingly.

To summarise the procedure used to generate a sequence of models and accretion rates, we remark that by analogy with stellar structure, for given core and envelope masses we have three state variables, $(P, T, R)$ to be determined as a function of $M(R)$ through equations \ref{eq:hydrostatic}--\ref{eq:lconv}.
We have boundary conditions for each of these at the surface and one for $R,$ at the surface of the core.
The problem thus appears to have one too many boundary conditions and thus be overdetermined.
This is resolved by allowing $L_{\rm tot}$ to be determined so that all the conditions can be satisfied.
Thus specification of the total mass together with the local disc properties specifies $L_{\rm tot}$.
This is then used to determine the accretion rate using equation \ref{eq:dmpdtequation} which in turn determines the rate at which the  envelope mass increases.

\section{Calculation of  the evolution of envelopes with free surfaces}
\label{sec:freecalc}

Once the protoplanetary disc has dispersed, the surface boundary conditions have to be changed 
from those used above so as to represent a free surface. The energy radiated into space is
as before supplied by gravitational settling but at fixed envelope mass which requires the radius to contract.
To obtain models belonging to   the contracting sequence, we proceed as above but implement  the different boundary conditions.
Firstly the radius $R_{\rm surf}$ is not fixed to be the radius of the Hill sphere but is determined in the course of the solution.
As for stellar models we take the radius to define the photosphere  where the optical depth is taken to be $2/3.$  
At this location the pressure is given by
\begin{equation}
P\kappa=g.
\end{equation}
The temperature is given by
\begin{equation}
T_{\rm surf} = (T_{*}^4+T^4_{\rm effp})^{1/4},
\end{equation}
where $T^4_{\rm effp} = L_{\rm tot}/(4\pi\sigma R^2_{\rm surf})$ with $L_{\rm tot}$ denoting the total luminosity escaping from the surface, and 
$T^4_{*} = L_{*}/(16\pi\sigma R^2_{p})$ with $L_{*}$ denoting the luminosity escaping from the central star.
Here $T_{\rm effp}$ is the effective temperature associated with the luminosity of the planet
and $T_{*}$ corresponds to the equilibrium temperature of a black sphere in the external radiation field of the star.
Thus we assume that  reprocessing of this radiation leads to a heating of the atmosphere that can be approximated as being spherically symmetric.
The interior boundary conditions at the surface of the core are as above.

The calculation of the models proceeds as above except that instead of the radius being fixed the envelope mass is.
Thus instead of obtaining a sequence for which the envelope mass is a function of the luminosity,
we obtain one for which the radius is a function of luminosity.
In this way we get  a sequence of models with fixed $M_{\rm p}$ and varying $R_{\rm surf}$,
each model being associated with a determined luminosity.
 
An evolutionary sequence of contracting models can  be determined by
equating the rate of increase of binding energy with the rate of radiation of luminous energy.
To do this the total energy of the envelope for a given model is found using equation \ref{eq:envenergy}.
Its derivative with respect to $R_{\rm surf}$ is found numerically from considering two models with slightly different $R_{\rm surf}$.
Then the radius can be evolved forwards in time having obtained  $dR_{\rm surf}/dt$ from
\begin{equation}
\label{eq:dmpdtequationfree}
\left(dE/dR_{\rm surf}\right) \left(dR_{\rm surf}/dt\right)=-L_{\rm tot}.
\end{equation}

\begin{figure*}
\centering
\includegraphics[width=0.32\textwidth]{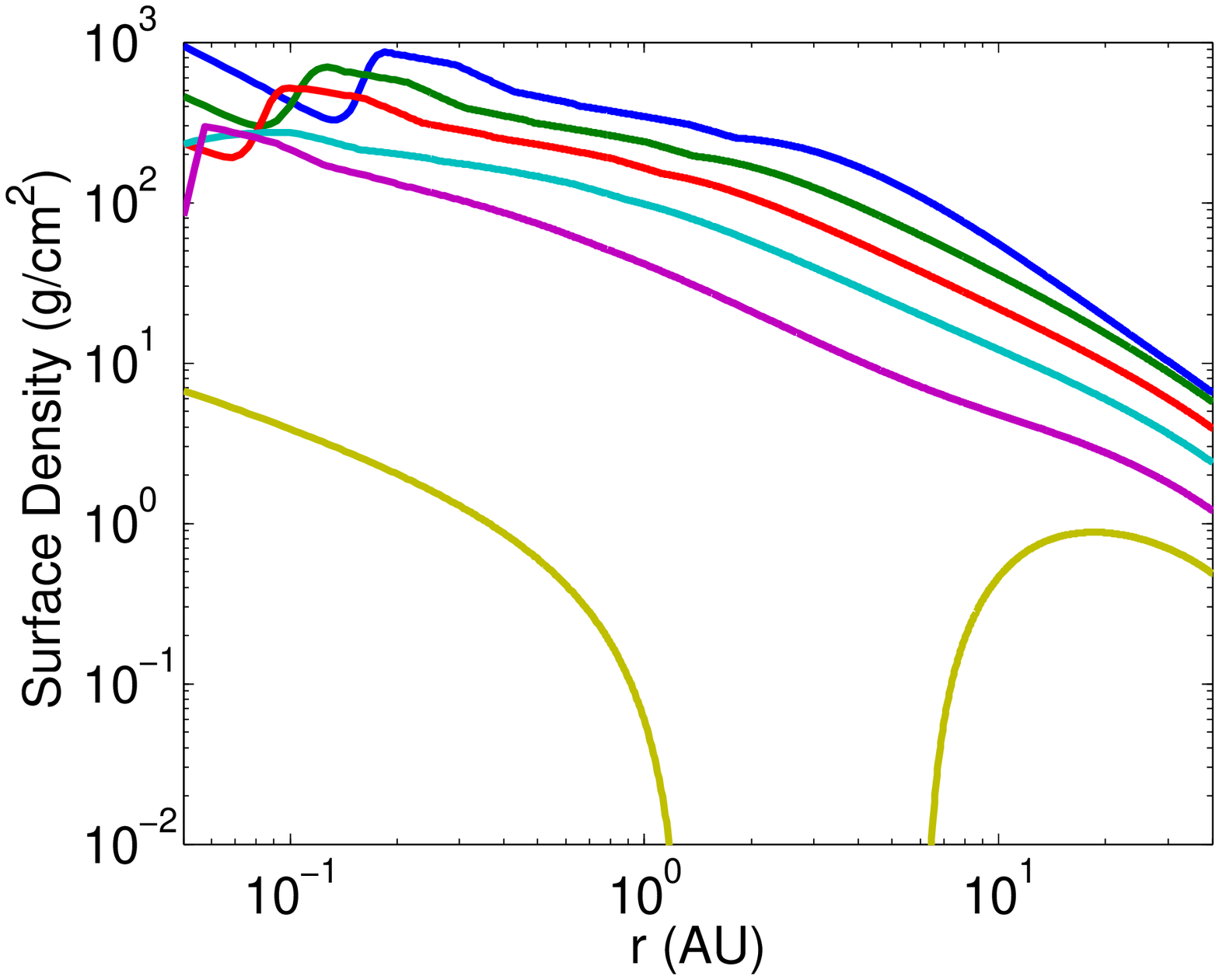}
\includegraphics[width=0.32\textwidth]{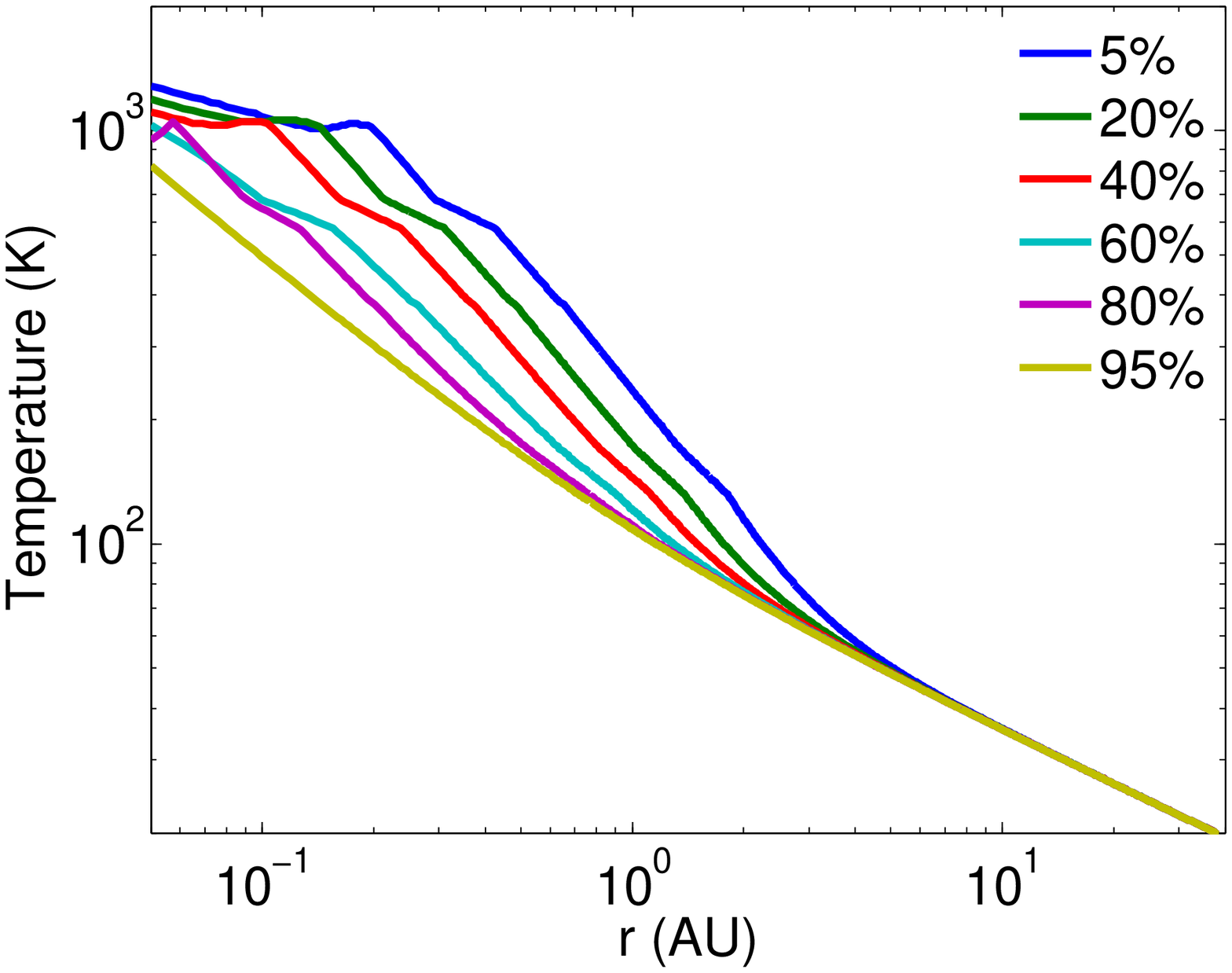}
\includegraphics[width=0.32\textwidth]{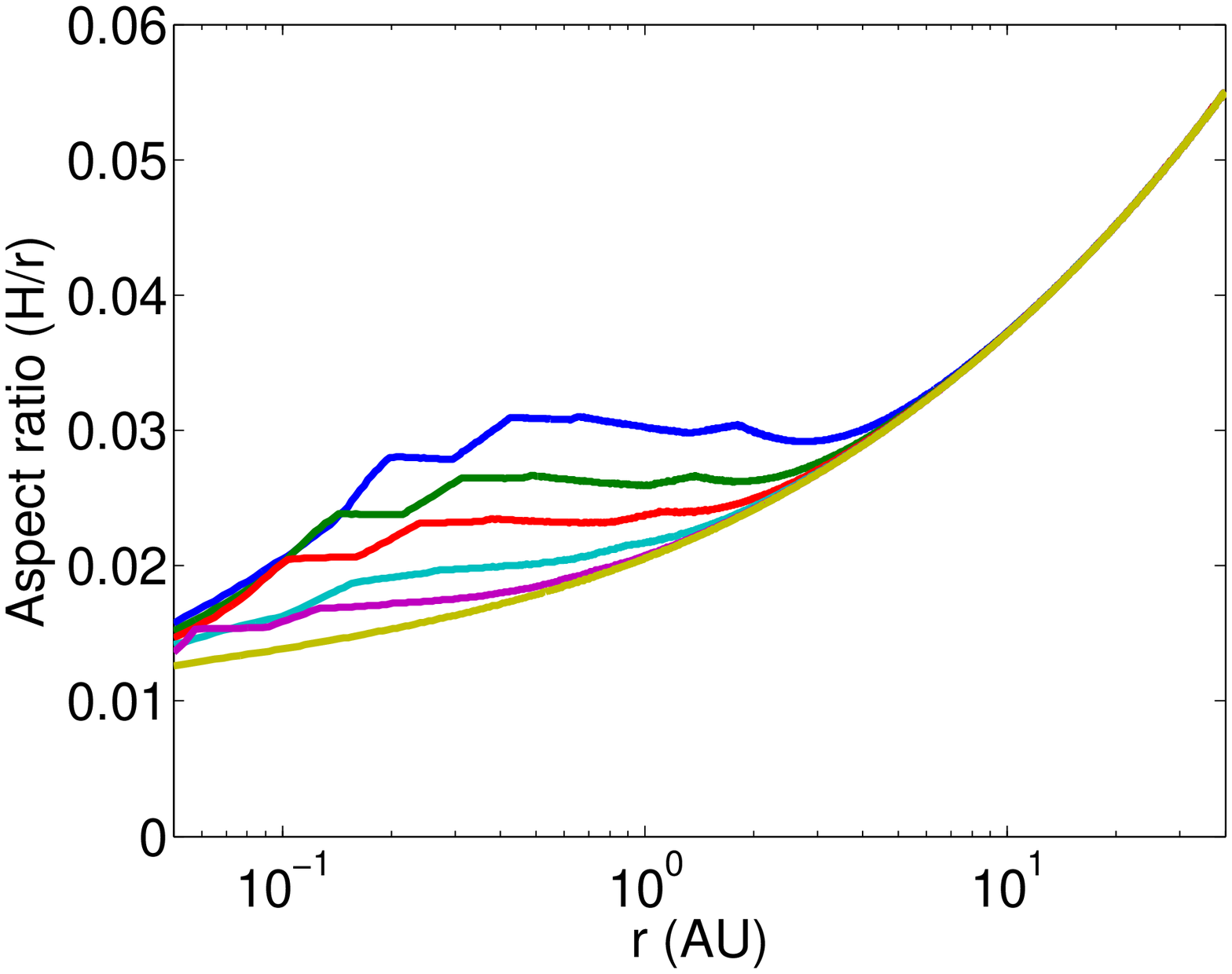}
\caption[Gas surface densities, temperatures and aspect ratios of an evolving disc]{Gas surface densities, temperatures and aspect ratios for 5, 20, 40, 60, 80 and 95 per cent (top-bottom lines)
of the disc lifetime in a $1 \times \mmsn$ disc (lifetime: 4.6 Myr)}
\label{fig:discevolution}
\end{figure*}

\subsection{End states}
\label{sec:endstatec}
The sequence of models described above have decreasing luminosity as they contract.
The envelope thus becomes denser and cooler.
As the only energy source considered is that due to gravitational settling of the envelope, the end state will be an isothermal envelope heated by stellar radiation with zero internal luminosity.
Attainment of this state could of course be retarded if a cooling core provides a heat source.
Accordingly, the lifetimes of inflated states might be underestimated by our calculations as we neglect this effect.

As the envelopes become cool and dense we found that in some cases the physical parameters
could become out of the range of the applicability of the equation of state of \citet{Saumon95}.
Accordingly our end state models were calculated  using the equation of state of 
\citet{Becker2014}. Tests showed that very similar results were obtained for models
using the different equations of state when the parameters were in the range of validity of both.

\section{Background disc model}
\label{sec:discs}
We utilise the disc model presented in \citet{ColemanNelson14,ColemanNelson16} where the standard diffusion equation for a 1D viscous $\alpha$-disc model is solved \citep{Shak,Lynden-BellPringle1974}. 
Temperatures are calculated by balancing black-body cooling against viscous heating and stellar irradiation. The viscous stress parameter $\alpha=2 \times 10^{-3}$ throughout most of the disc, but increases to $\alpha=0.01$ in regions where $T \ge 1000$ K to mimic the fact that fully developed magnetohydrodynamic turbulence can develop in regions where the temperature exceeds this value \citep{DeschTurner2015}.

\begin{table}
\centering
\begin{tabular}{lc}
\hline
Parameter & Value\\
\hline
Disc inner boundary & 0.05 $\au$\\
Disc outer boundary & 40 $\au$\\
$\Sigma_{\rm g}$(1 $\au$) & $1731 {\rm g/cm}^{-2}$\\
Metallicity & Solar\\
$\alpha$ & $2 \times 10^{-3}$\\
$\alpha_{\rm active}$ & $1 \times 10^{-2}$\\
Stellar Mass & $1\rm M_{\bigodot}$\\
$\rm R_{\rm S}$ & $2 \rm R_{\bigodot}$\\
$\rm T_{\rm S}$ & 4280 K\\
$f_{41}$ & 10\\
\hline
\end{tabular}
\caption{Disc and stellar model parameters. Note that the value for $\Sigma_{\rm g}$ applies only to the $1 \times \mmsn$ model and should be scaled appropriately for the 3 and $5 \times \mmsn$ models.}
\label{tab:discparameters}
\end{table}

The final stages of disc removal occur through a photoevaporative wind. A standard photoevaporation model is used for most of the disc evolution \citep[see][for a review]{Dullemond}, corresponding to a photoevaporative wind being launched from the upper and lower disc surfaces. Direct photoevaporation of the disc is switched on during the final evolution phases when an inner cavity forms in the disc (due to the inner parts viscously accreting onto the star), corresponding to the outer edge of the disc cavity being exposed to the stellar radiation \citep{Alexander09}. We note that when a giant planet begins to form in a disc, and gap opening criteria are met, then tidal torques from the planet are able to open a gap using the impulse approximation from \cite{LinPapaloizou86}. 

Figure \ref{fig:discevolution} shows the evolution of a 1 $\times \mmsn$ \citep[Minimum Mass Solar Nebula,][]{Hayashi} disc model.
Disc surface density profiles are shown in the left-hand panel, temperature profiles are shown in the middle panel, and $H/r$ profiles are shown in the right-hand panel. The times corresponding to each profile are shown in the middle panel, expressed as a percentage of the disc lifetime ($\sim$ 4.6 Myr for a 1 $\times \mmsn$ disc).
The active turbulent region where the local disc temperature exceeds 1000 K can be seen in the inner region of the disc, gradually moving in towards the star as the surface density decreases, reducing the viscous heating rate and the opacity.
This region disappears when the temperature no longer exceeds 1000 K anywhere in the disc, as occurs at the evolutionary stage illustrated by the yellow curves in Fig. \ref{fig:discevolution}.
This happens in all of the disc models when the gas surface density in the inner disc drops below $\sim$ 100 $\rm g/cm^2$, which typically occurs 0.5 Myr before complete dispersal of the gas disc.

On longer time-scales the removal of gas by the photoevaporative wind causes the disc to disperse.
The loss of mass at large radius results in the inner disc emptying viscously onto the star leaving a gap that separates the remnant inner disc from the outer disc which is clearly seen in the surface density profile at the evolutionary stage illustrated with yellow curves in Fig. \ref{fig:discevolution}.
Finally the remnant outer disc is removed by the wind \citep{Clarke2001}.

When the disc mass is increased to larger values, as will be considered in this work, the evolution is qualitatively similar to that of the 1 $\times \mmsn$ disc.
The major difference is in the lifetime of the disc.
We find that 1 $\times \mmsn$ discs have lifetimes $\sim 4.6$ Myr, whereas 3 and 5 $\times \mmsn$ discs last for $\sim$ 7.8 and 9.4 Myr respectively.
Apart from existing for longer, the main effect that increasing the disc mass will have on gas accreting planetary cores is to change the physical conditions at their outer boundaries where their gaseous envelopes merge with the gas disc.
For more massive discs there will be an increase in the viscous heating rate, leading to slightly higher disc temperatures, and the local gas density and pressure will also be increased.
These increases will affect the structure of the gaseous envelope and modify the gas accretion rate. In practice, however, we find that cores of a given mass located at a particular orbital radius evolve very similarly in the different disc models until they reach a critical state, because the local conditions in the models do not vary by a large amount.
Once the planets have reached a critical state, they undergo runaway gas accretion and then begin to accrete gas from the disc at the viscous supply rate. The evolution of these planets from this point then depends on the remaining disc mass and lifetime, which will lead to different evolution paths for similar planets embedded in discs with different masses.

Table \ref{tab:discparameters} shows the disc parameters that are used in all of the simulations. The value of $f_{41}$ defines the value of the photoevaporative UV flux that is incident on the disc, as described by \cite{Dullemond}.

\section{In situ accretion of gaseous envelopes}
\label{sec:insitu}
We now present one fiducial model in detail, before discussing the results for cores of different masses that are placed at different locations in the three disc models that we consider.

\begin{figure*}
\centering
\includegraphics[width=0.46\textwidth]{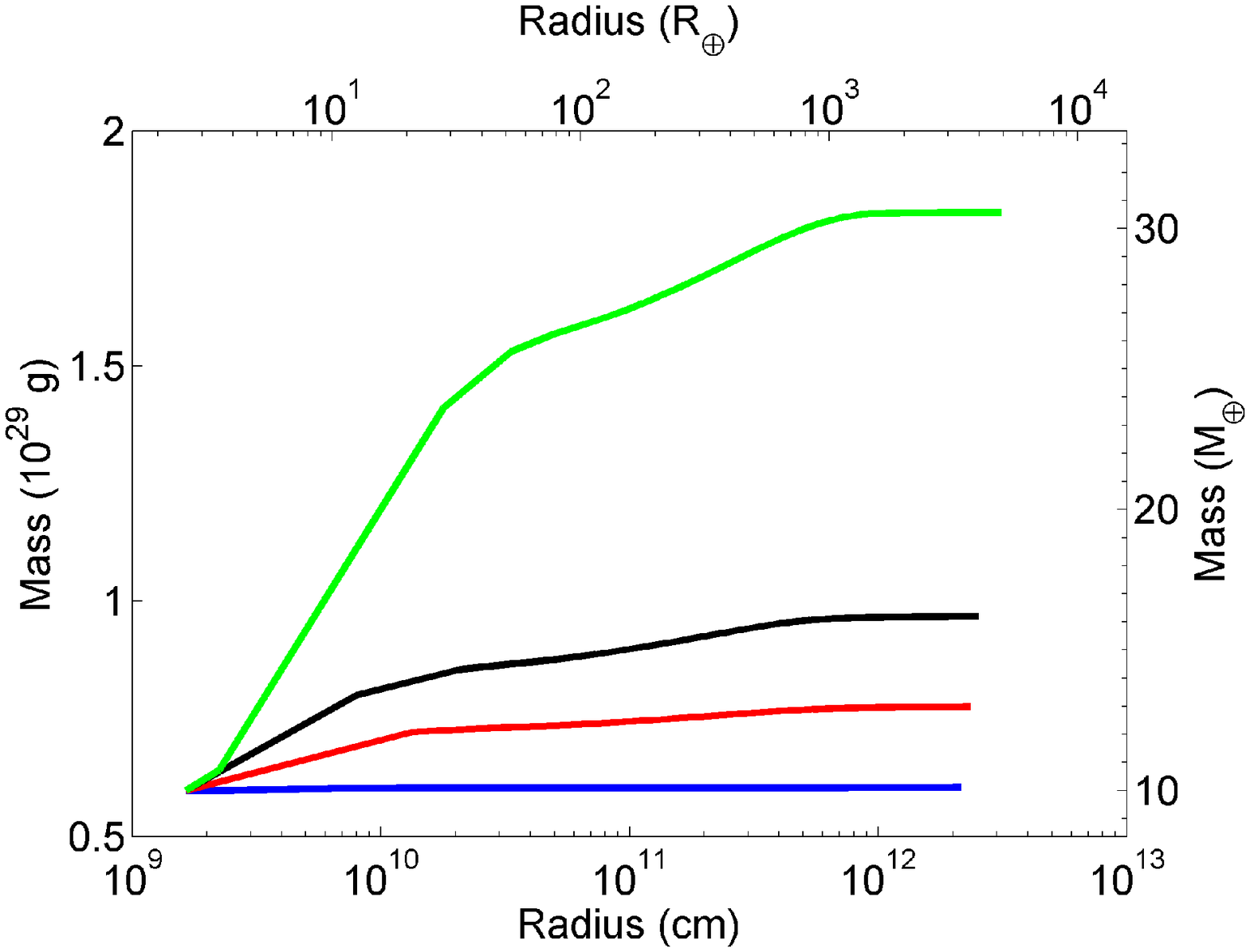}
\includegraphics[width=0.46\textwidth]{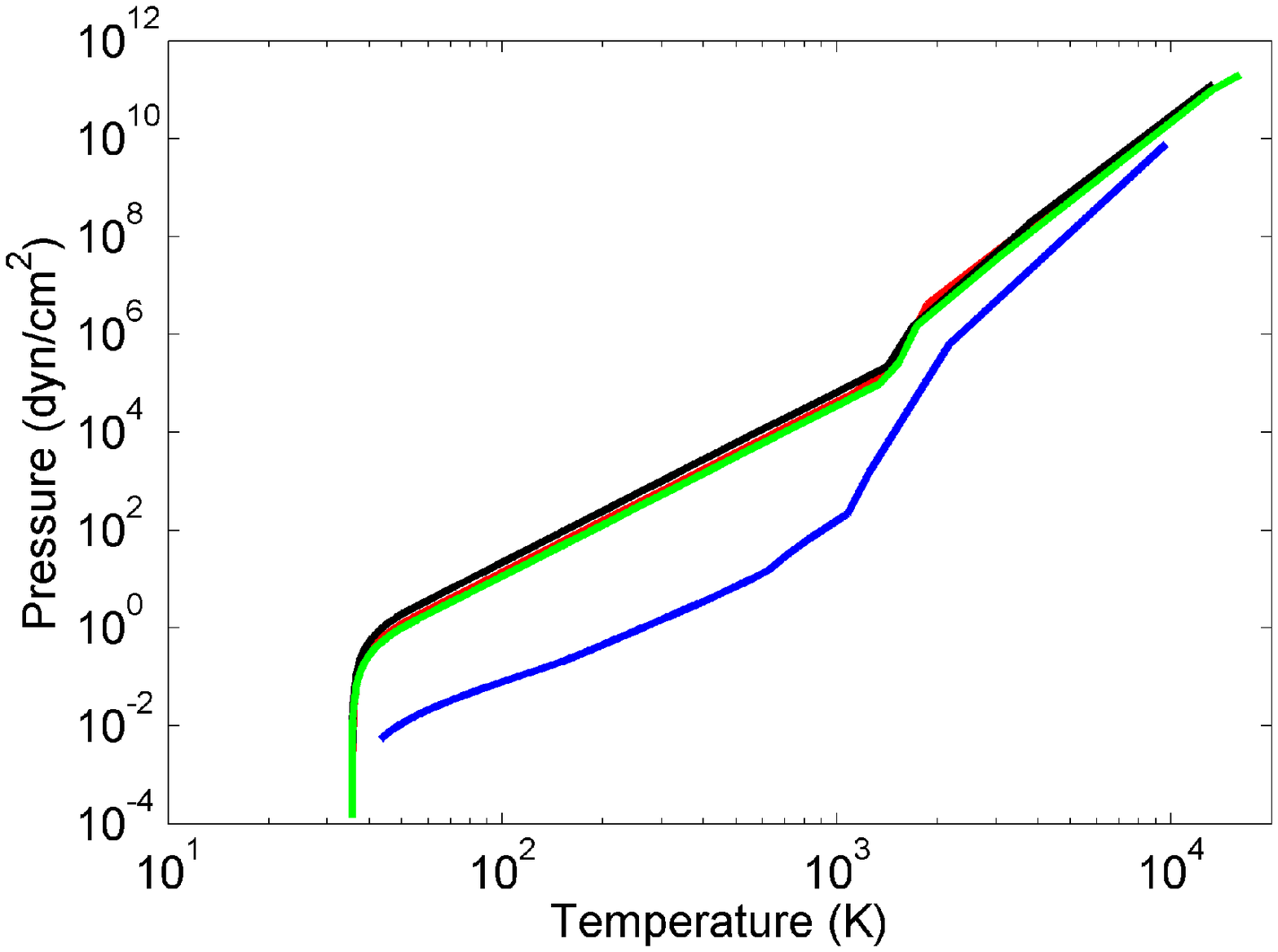}\\
\includegraphics[width=0.46\textwidth]{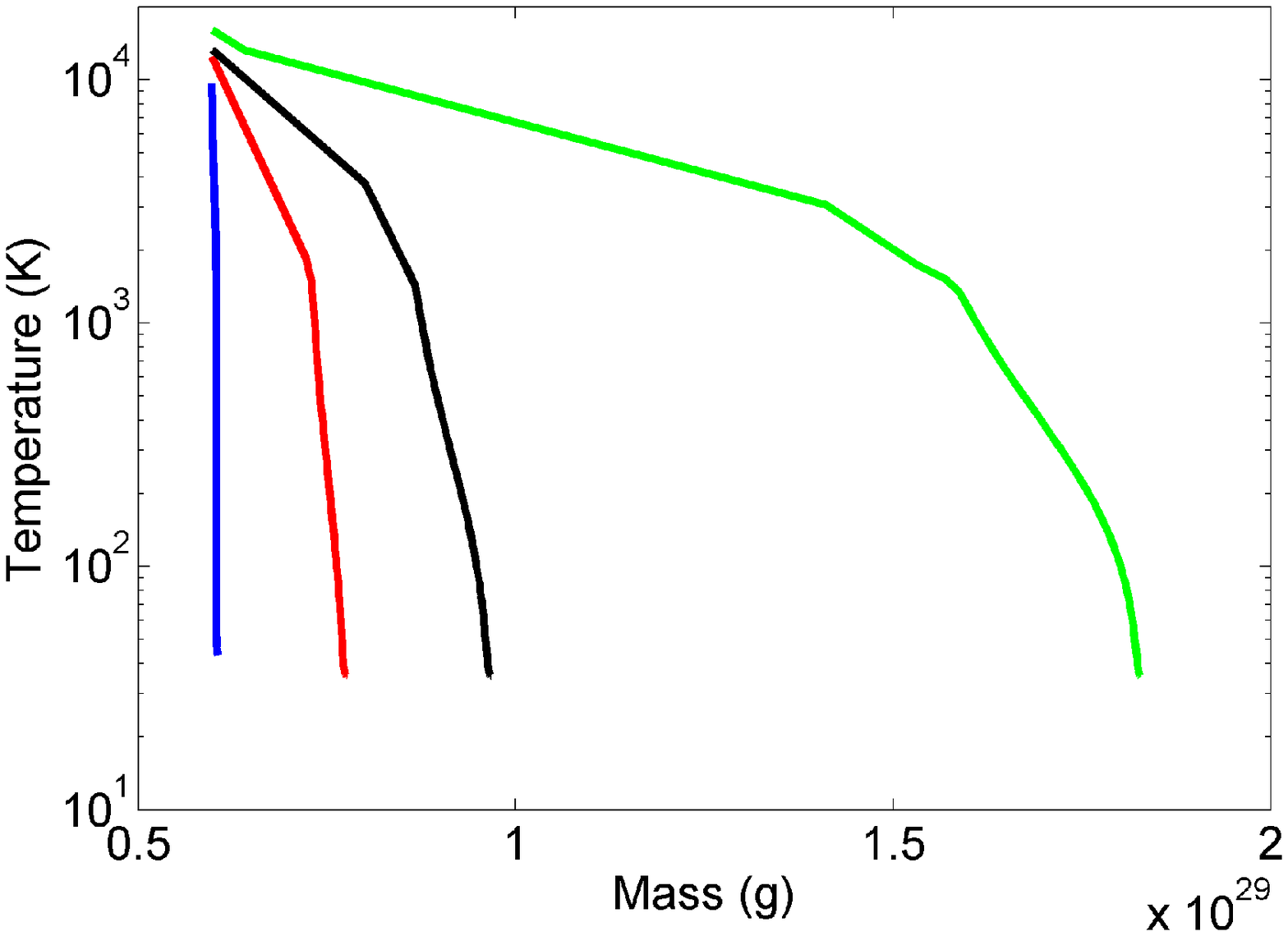}
\includegraphics[width=0.46\textwidth]{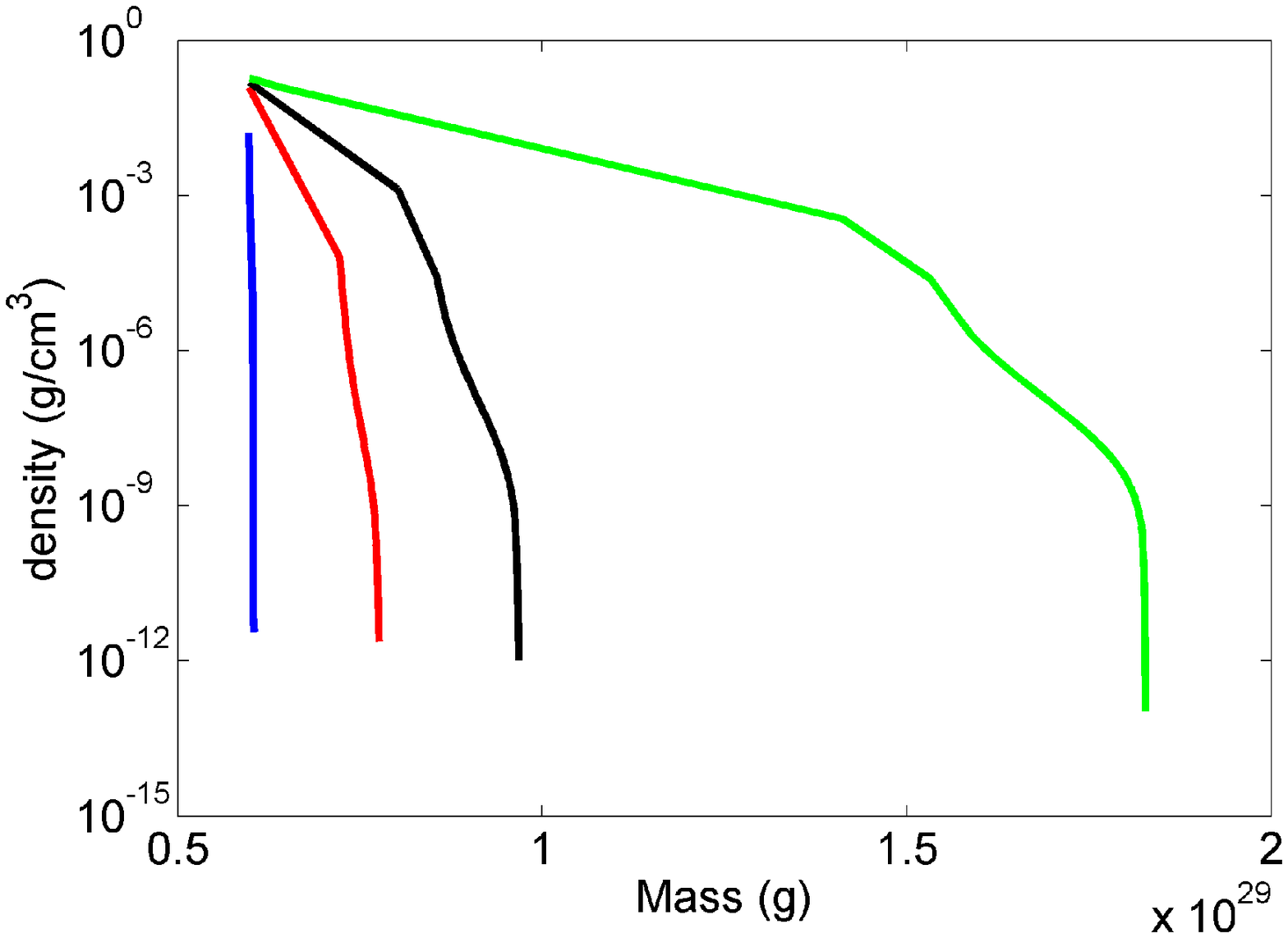}
\caption{Envelope structure for a 10 $\me$ core located at 10 $\au$ in a 1 $\times \mmsn$ disc. The top left-hand panel shows mass as a function of radius. The top right-hand panel shows pressure as a function of temperature. The bottom left-hand panel shows temperature as a function of mass. The bottom right-hand panel shows density as a function of mass. The different coloured lines show different times of evolution: blue -- 0.1 Myr, red -- 1 Myr, black -- 2.5 Myr, and green -- 4.2 Myr.}
\label{fig:atmos_example}
\end{figure*}

\subsection{Envelope evolution}
\label{sec:envelope_evolution}
We solve equations \ref{eq:hydrostatic}--\ref{eq:lconv} using the boundary conditions described in Section \ref{sec:boundary} to obtain the envelope structure models.
In this work we only consider single planets that are embedded in protoplanetary discs. We use core masses of 5, 10 and 15 $\me$, and we place them in discs of mass 1, 3 and 5 $\times \mmsn$. The planets are placed at different locations in the disc: 0.1, 0.2, 0.5, 1, 2, 5 and 10 $\au$. We embed the planets at different locations in the disc to examine the effects that the different disc locations and conditions have on the accretion of gaseous envelopes.
The combination of these parameters results in a total of 63 different simulations. 
To allow the disc time to settle into a quasi-steady state, we inject the cores into the disc after a time of 0.1 Myr. From this point onwards the cores are able to accrete gas from the disc following equation \ref{eq:dmpdtequation}, so that the rate of liberation of accretion energy balances the planet's luminosity. All gas that is accreted by the planet is removed from the surrounding disc.

When calculating a sequence of hydrostatic models for which the mass of the gaseous envelope increases, one may reach a point where a hydrostatic model can no longer be obtained. Here, the luminosity associated with the liberation of gravitational energy by the settling envelope is insufficient to allow the envelope to be supported, and it is expected that the planet will then enter a period of runaway gas accretion. In this paper we refer to this moment as being when the envelope reaches a critical state. If the planet's envelope reaches a critical state, we change the gas accretion routine to that used in \citet{ColemanNelson16b}, which utilises fits based on models from \citet{Movs} to calculate the gas accretion rate, including during the early stages of runaway accretion. In our models, once in this phase of runaway gas accretion, the planet rapidly accretes material from within its feeding zone. Once the material in the feeding zone has been accreted, the planet opens up a gap in the protoplanetary disc as indicated by the torque formulae of \citet{LinPapaloizou86}. At this point the planet starts to accrete gas at the viscous supply rate. If the protoplanetary disc disperses before the planet's envelope reaches a critical state, however, we transition the envelope to a free boundary model and allow it to cool and settle over time as discussed in Section~\ref{sec:lte}.

We will now present the results of the simulations. To begin, we will examine a single case where we show the envelope structure and present its evolution over time. We will then examine the effects that placing the planetary cores in different regions of the discs has on the accretion and evolution of the envelopes, before considering the effects that different initial core masses have on their evolution.

\subsection{A fiducial envelope accretion calculation}
\begin{figure*}
\centering
\includegraphics[width=0.46\textwidth]{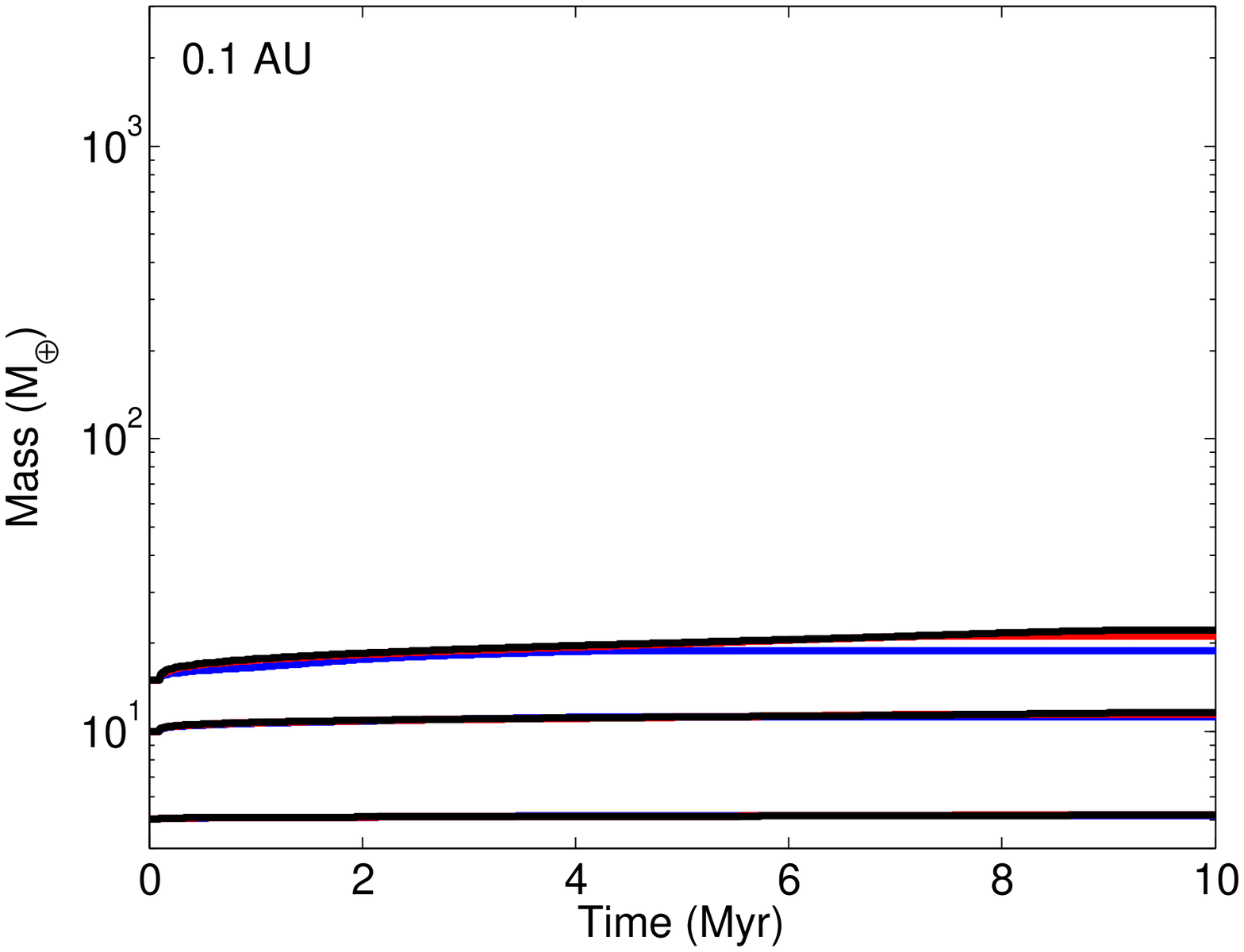}
\includegraphics[width=0.46\textwidth]{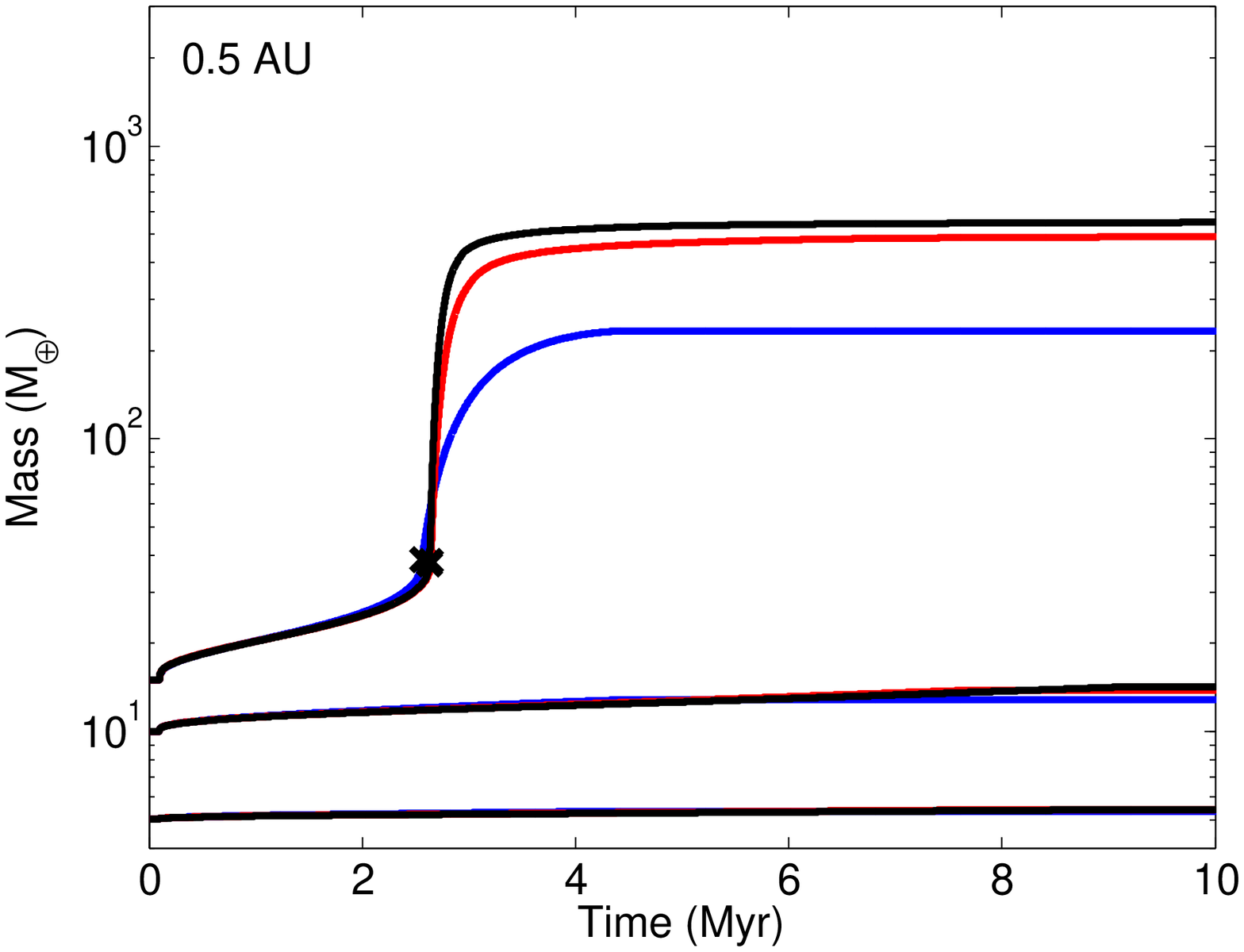}\\
\includegraphics[width=0.46\textwidth]{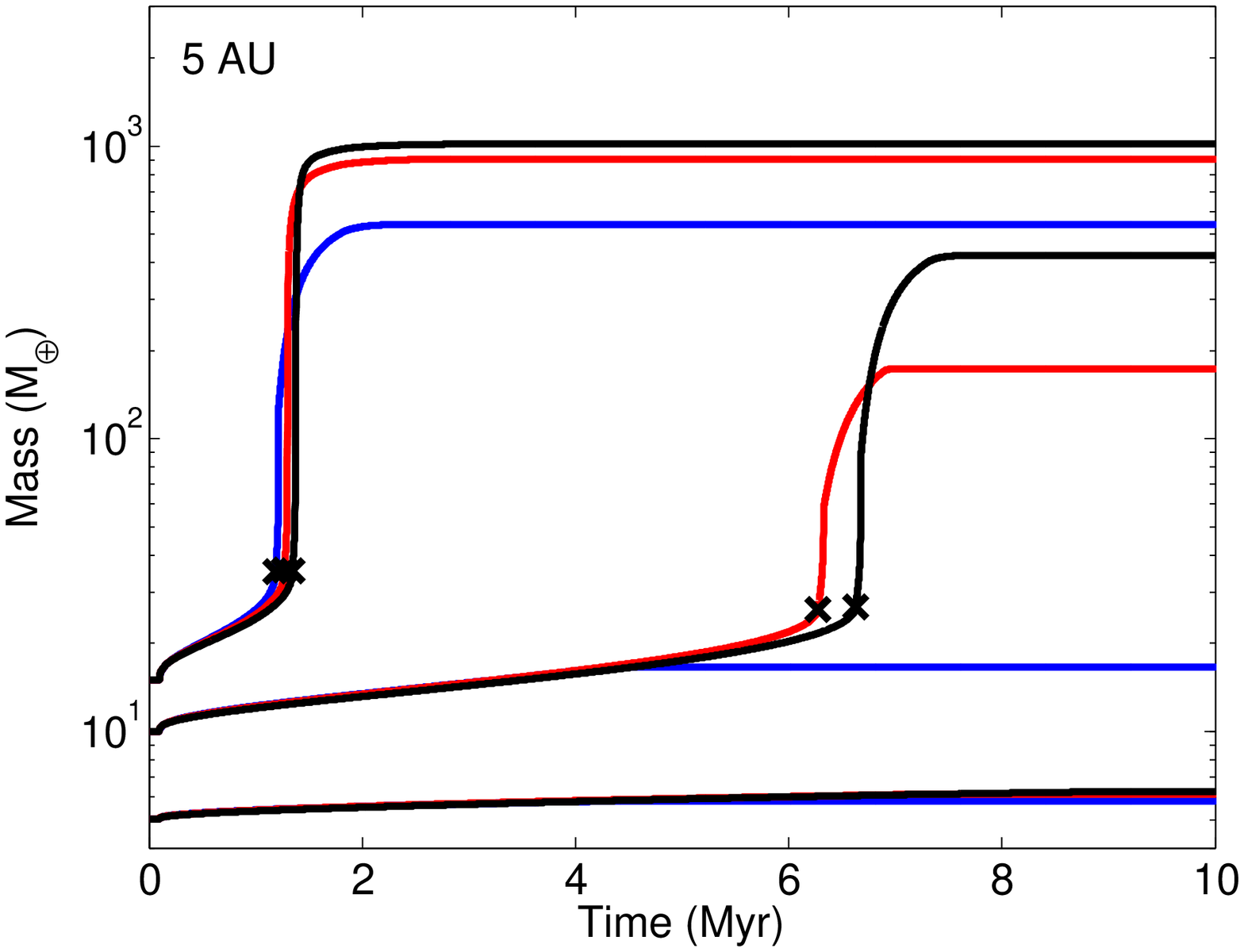}
\includegraphics[width=0.46\textwidth]{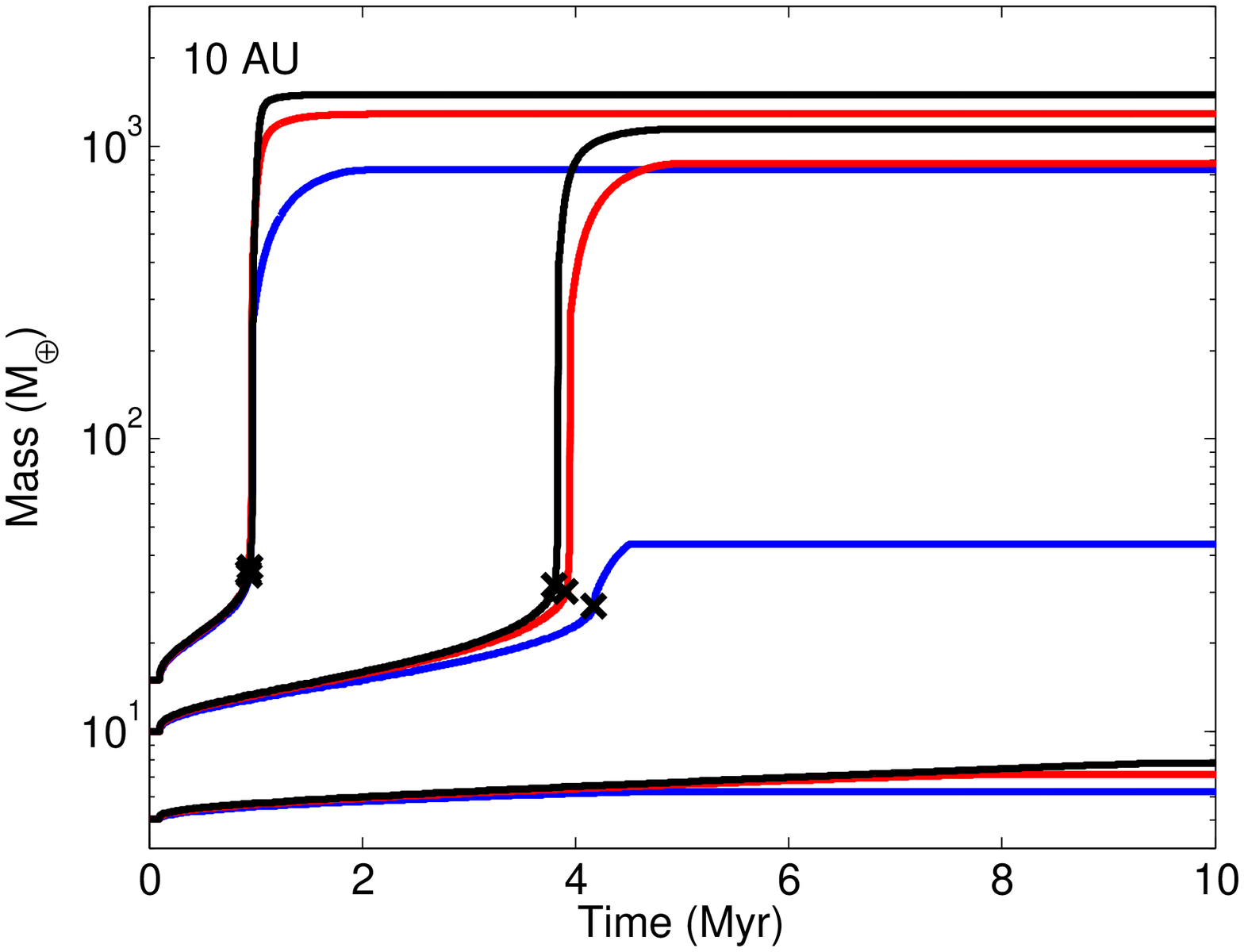}
\caption{Mass evolution for 5, 10 and 15 $\me$ cores accreting gaseous envelopes at 0.1 $\au$ (top left), 0.5 $\au$ (top right), 5 $\au$ (bottom left) and 10 $\au$ (bottom right). The different coloured lines represent different initial disc masses: blue = 1 $\times \mmsn$, red = 3 $\times \mmsn$, and black = 5 $\times \mmsn$. The black crosses show when a planet's envelope has reached a critical state.}
\label{fig:sma_evolve}
\end{figure*}

\begin{figure*}
\centering
\includegraphics[width=0.3\textwidth]{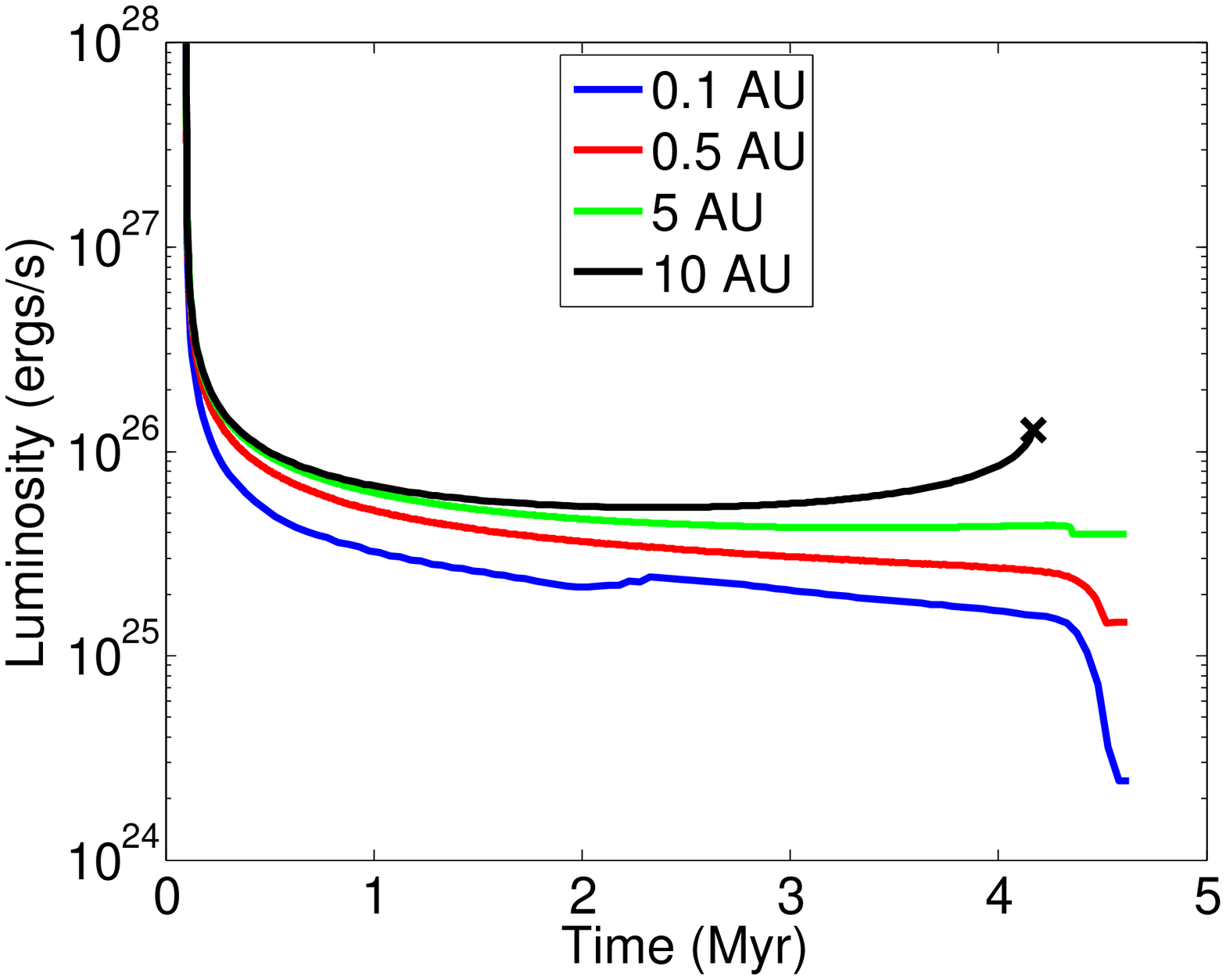}
\includegraphics[width=0.3\textwidth]{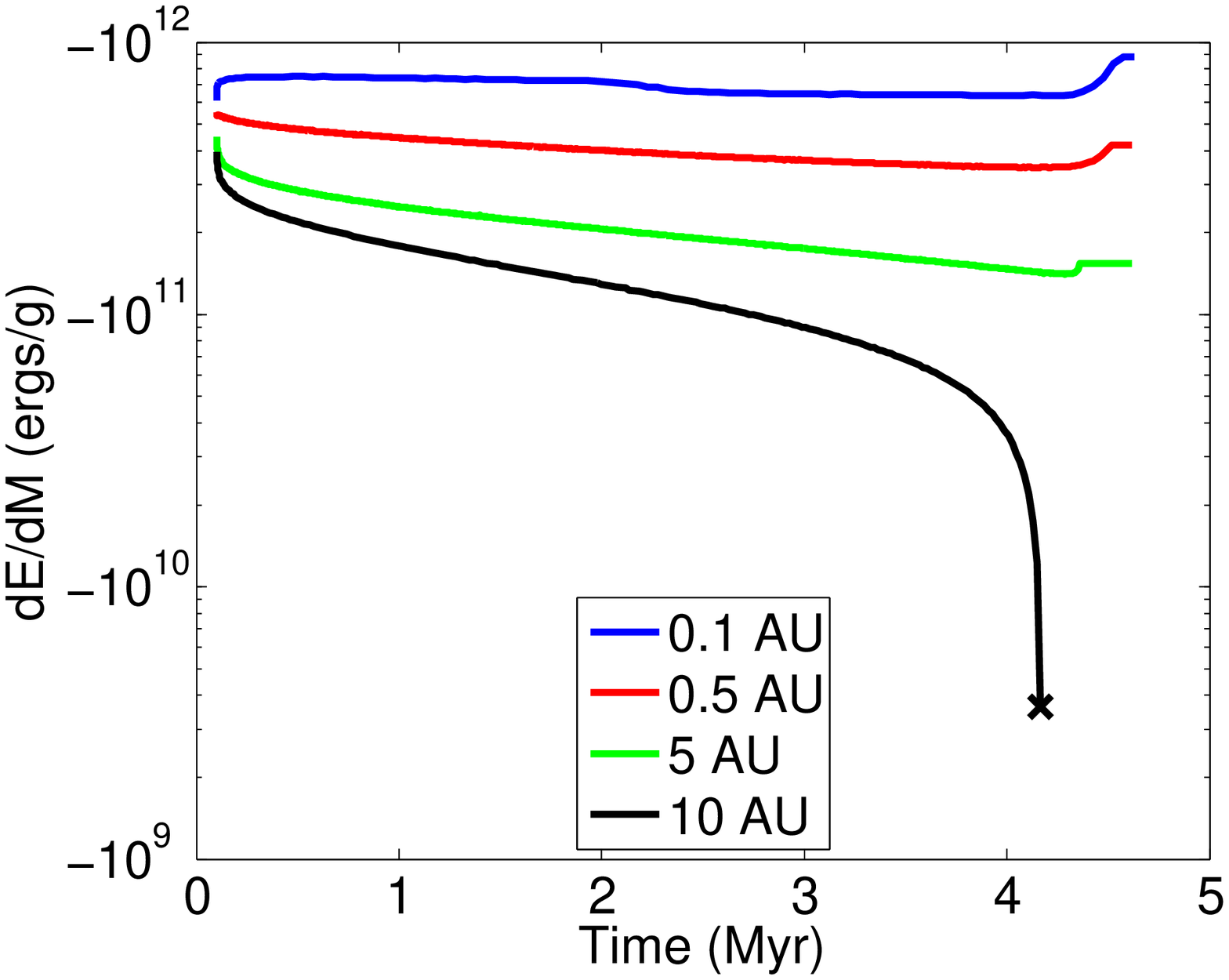}
\includegraphics[width=0.3\textwidth]{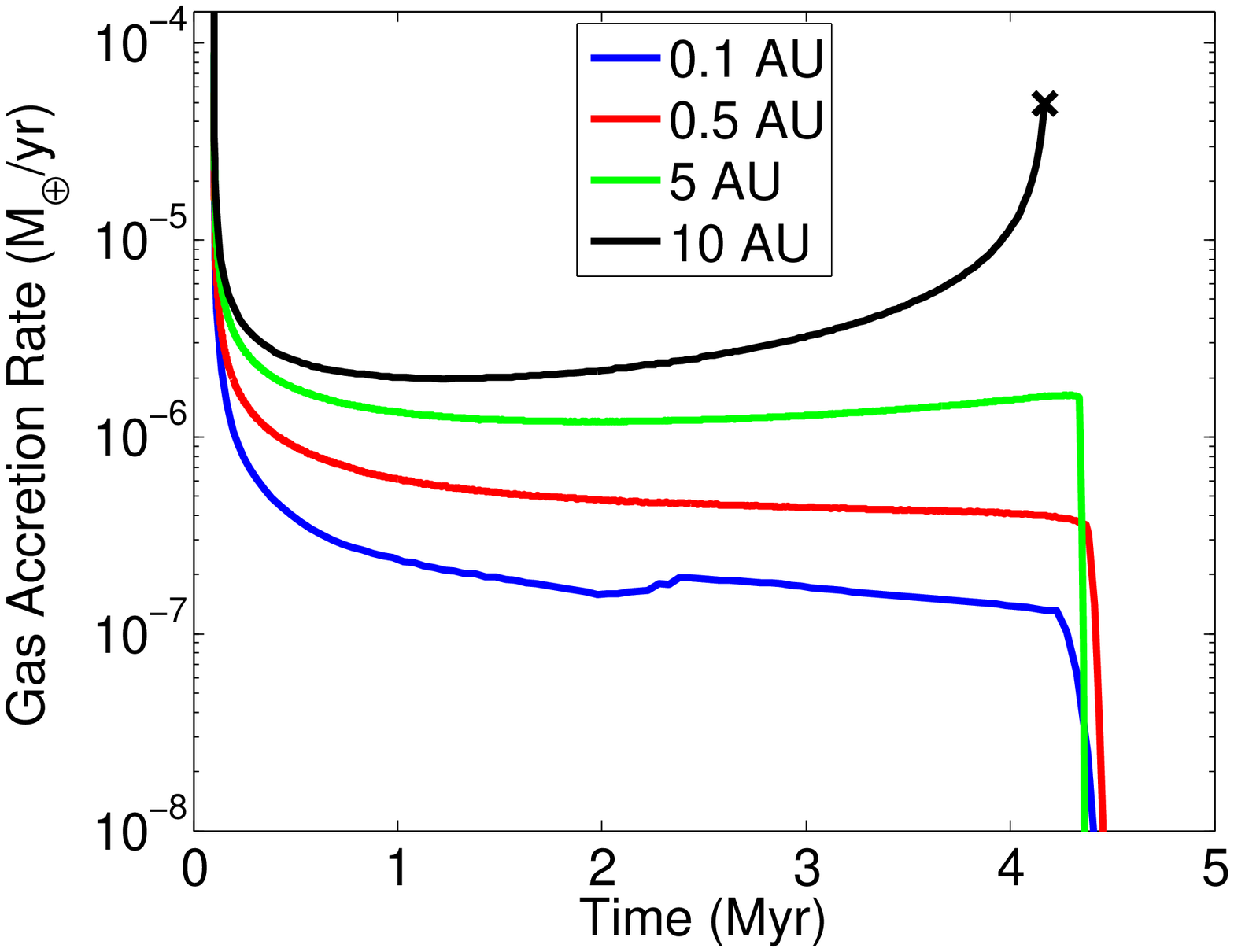}
\caption{Luminosity evolution (left panel), Energy released per unit mass through settling of the envelope (middle panel), and gas accretion rates (right panel) on to 10 $\me$ cores over time at different locations in a $1 \times \mmsn$ disc.
The black crosses represent where a planet's envelope has reached a critical state.}
\label{fig:time_gas_acc}
\end{figure*}

We begin by describing the time evolution of a gaseous envelope accreting onto a 10 $\me$ core located at 10 $\au$ in a 1 $\times \mmsn$ disc. We note that the radius of the core is $1.7 \times 10^9$ cm.
Figure \ref{fig:atmos_example} shows the internal structure of this envelope at four moments in time: blue line -- 0.1 Myr; red line -- 1 Myr; black line -- 2.5 Myr; green line -- 4.2 Myr. The temporal evolution of the mass of this envelope is shown by the blue line that starts at $10 \me$ in the lower right panel of Fig. \ref{fig:sma_evolve}. The evolution of the luminosity, energy liberated per unit mass, and the mass accretion rate are shown by the black lines in Fig. \ref{fig:time_gas_acc}.

We initialise the calculations with an envelope mass that is equal to 1 \% of the core mass, i.e. a 0.1 $\me$ envelope is assumed to be present initially surrounding the 10 $\me$ core. The model then calculates the planet luminosity that is required to support this envelope according to the assumed equation of state, opacity and boundary conditions. The blue lines in Fig. \ref{fig:atmos_example} show the structure of this initial envelope. As expected, the innermost part of the atmosphere is convective. This convection zone extends from the core surface to $\sim 10^{10}$ cm, and contains $\sim 90 \%$ of the envelope's mass.
There is also a second convection zone that extends between 7 $\times 10^{10}$ cm to $10^{12}$ cm. These convection zones occur for temperatures $T>1860$ K and $55$ K $\le T \le 570$ K, respectively.

As time progresses, the gaseous envelope settles onto the planet's core as more material is accreted from the surrounding disc. One effect of this is to cause the luminosity of the planet to decrease as the optical depth through the atmosphere increases, as shown in the left panel of Fig. \ref{fig:time_gas_acc}. After 1 Myr, shown by the red lines in Fig. \ref{fig:atmos_example}, the planet has a mass of 13 $\me$, with the gaseous envelope containing $3 \me$. There are two convection zones, one in the deep interior of the planet extending out to $1.2 \times 10^{10}$ cm, and one further out between the radii $3 \times 10^{10}$ cm and $5 \times 10^{11}$ cm.
These two convection zones differ from those in the initial envelope because of the temperatures at which they occur. Compression has heated the inner envelope, and the temperatures in the innermost convective zone now exceed T $> 2340$ K. The outer convection zone occurs for $55$ K $\le T \le 1280$ K. At this time, approximately 66 \% of the envelope's mass lies within the inner convection zone, with a further 21 \% being situated within the outer convective zone.

The black lines in Fig. \ref{fig:atmos_example} show the envelope structure after 2.5 Myr. Up until this point, the luminosity of the planet has been decreasing due to the increasing opaqueness of the envelope arising from its early compression (see Fig. \ref{fig:time_gas_acc}). After 2.5 Myr, however, the luminosity and gas accretion rates start to increase as the self-gravity of the envelope helps to compress it. Here the gaseous envelope has a mass of 6.2 $\me$, and hence comprises $\sim 40 \%$ of the total planet mass.
The inner convective zone again contains $\sim 66 \%$ of the envelopes mass, and extends out to 1.7 $\times 10^{10}$ cm.
The outer convective zone extends between $3.4 \times 10^{10}$ cm and $5.1 \times 10^{11}$ cm and contains $\sim 25\%$ of the envelope's mass.
The temperatures in the outer and inner convection zones are $55$ K $\le T \le 1400$ K and T $> 2160$ K, respectively.

After a further 1.7 Myr, corresponding to a total of 4.2 Myr of evolution, the planet reaches a critical state and is about to enter runaway gas accretion.
The green lines in Fig. \ref{fig:atmos_example} show the envelope structure at this point.
The planet has a total mass of 30.6 $\me$ with the envelope contributing $67 \%$ of this.
The inner convective zone now extends out to $\sim 2.8 \times 10^{10}$ cm and contains $\sim 73 \%$ of the envelope's mass, whilst the outer convective zone lies between $6 \times 10^{10}$ cm and $7.7 \times 10^{11}$ cm and contains $\sim 20 \%$ of the envelope's mass. These convective zones occur at similar temperatures to those found for the convective zones of the atmosphere after 2.5 Myr. 

We recall that the disc model in which the planet is embedded disperses after 4.6 Myr. As described in Sect.~\ref{sec:envelope_evolution}, when the planet envelope reaches the critical state then the assumption of hydrostatic equilibrium breaks down and another prescription for mass evolution of the planet is required. Our adoption of fits to the mass versus time plots presented in \citet{Movs}, followed by assuming that the planet grows at the viscous supply rate of gas by the protoplanetary disc when the planet becomes gap forming, do not allow us to calculate the detailed structure of the planet, so the green lines in Fig. \ref{fig:atmos_example} represent the final time at which the detailed envelope structure can be computed. The longer term mass growth of the planet, however, is shown in the bottom right panel of Fig. \ref{fig:sma_evolve}, which is discussed in more detail in the next section.

\subsection{Effect of location}
As shown in Fig. \ref{fig:discevolution}, the disc models that we employ have strongly varying temperatures as a function of radius, and this variation in temperature has a significant influence on the accretion histories of the planets that are placed at different locations in the discs, a point that has been explored in previous work \citep[e.g.][]{Piso2014}. For a given time interval over which accretion occurs, gaseous envelopes that accumulate onto planets close to the central star are expected to be less massive than those accreted by planets at large orbital radii. The gas orbiting close to the star is much hotter than that further away, reducing gas accretion rates because of its higher thermal energy content. The colder gas located at large orbital radii is accreted more efficiently because its lower thermal energy allows it to become more deeply bound to the accreting cores. From the point of view of constructing a sequence of hydrostatic models, as undertaken here, increasing the disc temperature increases the surface temperatures of the models through equation \ref{eq:Tsurf}, and this  in turn increases the temperatures and entropies throughout the envelopes, leading to longer cooling times and slower accretion rates.

Figure \ref{fig:sma_evolve} shows the mass evolution of 5, 10 and 15 $\me$ cores situated at 0.1 $\au$ (top left-hand panel), 0.5 $\au$ (top right-hand panel), 5 $\au$ (bottom left-hand panel) and 10 $\au$ (bottom right-hand panel) in their respective discs. The blue lines show the evolution in 1 $\times \mmsn$ discs, with the red and black lines denoting 3 and 5 $\times \mmsn$ discs respectively, and the black crosses indicate that a planet has reached a critical state and is about to undergo runaway gas accretion.

\subsubsection{Accretion at $a_{\rm p} \le 0.5 \au$}
The top left-hand panel of Fig. \ref{fig:sma_evolve} shows the evolution of the masses of the planets orbiting at 0.1 $\au$. The evolution of the luminosity, change in energy liberated per unit mass, and accretion rate for the planet with a $10 \me$ core located in a $1 \times \mmsn$ disc located at $0.1 \au$ is shown by the blue line in Fig. \ref{fig:time_gas_acc}. 
These planets are in a hot and dense environment, with temperatures moderately exceeding 1000 K for the majority of the disc lifetime (we note that dust grains remain an important source of opacity in the models at these temperatures).

It is clear from Fig. \ref{fig:sma_evolve} that none of the planets located at $0.1 \au$ accrete large amounts of gas and reach the critical state that precedes runaway gas accretion. As such, our models predict that none of these planets can grow to become gas giants. The most massive envelope that could be accreted was in a 5$\times \mmsn$ disc, where a 15 $\me$ core accreted a 7.2 $\me$ envelope, representing $\sim 33 \%$ of the planet's total mass.
The core was able to accrete this significant envelope due to the long lifetime of the disc ($\sim 9.4$ Myr), and the high core mass which enhances the gas accretion rate as discussed in Section \ref{sec:core-mass}. 
The $10 \me$ cores located at $0.1 \au$ were only able to accrete between $1.2 \me$ and $1.6 \me$, with more gas accreting onto cores in the more massive and longer lived disc models. The $5 \me$ cores accreted very small amounts of gas, with the maximum amount being $0.2 \me$ which accreted in the long-lived $5 \times \mmsn$ model.

For a given core mass, we find that the outcomes of the calculations for planets located at $0.1 \au$ are broadly similar when varying the disc model, and this is to a significant degree because in all models the disc temperature at this location is $\sim 1000$ K for a large part of the evolution, due to our inclusion of a viscosity switch at $T \ge 1000$ K to mimic the development of fully developed magnetised turbulence in the hot inner disc. As the disc model evolves, the location of the highly viscous region moves inwards, and it eventually moves interior to $0.1 \au$ at which point the planet finds itself in a cooler environment. This normally occurs approximately 2 Myr from the end of the disc lifetime in all models, and causes there to be a modest increase in the gas accretion rate as shown by the blue lines in Fig. \ref{fig:time_gas_acc} for the planet with a $10 \me$ core.

Moving the cores further out into cooler disc regions results in enhanced accretion rates, and we can see from the top right panel of Fig. \ref{fig:sma_evolve} that placing cores at $0.5 \au$ leads to the accumulation of more massive envelopes. In particular, the models with $15 \me$ cores all reach the critical state and undergo runaway gas accretion to become gas giants. For a given core mass, we can see that the planet evolution models are rather insensitive to the disc models. The planets with $15 \me$ cores all reach the critical state after  $\sim 2.6$ Myr when their total masses are $\sim 35$ - $40 \me$, with the gaseous envelopes contributing approximately $60\%$ of their mass. After reaching the critical state these planets undergo runaway gas accretion and become gas giants by rapidly accreting the gas from their feeding zones before accreting gas from the disc at the viscous rate. These planets are finally able to reach masses $0.8 M_{\rm J} \le m_{\rm p} \le 2 M_{\rm J}$.

Considering the planets with 10 $\me$ cores, it was found that they are unable to accrete enough gas to reach the critical state. In the $1 \times \mmsn$, with a lifetime of 4.6 Myr, the 10 $\me$ core was only able to accrete $\sim 2.8 \me$ of gas. The planet in the $5 \times \mmsn$ disc accreted a $\sim 4.3 \me$ gas envelope, comprising $30\%$ of the planet's total mass. This increase in gas accretion is due to the disc having a longer lifetime of 9.4 Myr.
The 5 $\me$ cores located at $0.5 \au$ were still unable to accrete significant gaseous envelopes on account of their low masses. Envelope masses for these cores ranged between 0.3 $\me$ and 0.4 $\me$, with the latter envelope being accreted in the more massive, longer lived disc.

To understand the evolution of the planets as a function of orbital radius, it is useful to consider Fig. \ref{fig:time_gas_acc}. We recall from equation \ref{eq:dmpdtequation} that the mass accretion rate of the planet is controlled by the luminosity of the hydrostatic model and the quantity $dE /dM_{\rm p}$, which measures the rate at which the total energy of the planet changes as a function of its mass. During the early evolution, the left panel in Fig. \ref{fig:time_gas_acc} shows that the total luminosity drops rapidly as the envelope contracts and the optical depth through the envelope increases, but thereafter the luminosity is approximately constant throughout most of the evolution. We also see that for models that do not undergo runaway gas accretion, $dE / dM_{\rm p}$ is also essentially constant during the evolution. So gas accretion occurs at an almost constant rate throughout most of the evolution. We note, however, that the luminosity required for the models to achieve hydrostatic balance is small at small orbital radii and increases for planets located further from the star, and $dE / dM_{\rm p}$ decreases in magnitude as one moves the models further from the star. This results in the gas accretion rate increasing with increasing stellar distance for a fixed core mass, as observed in the models.

\begin{figure*}
\centering
\includegraphics[width=0.3\textwidth]{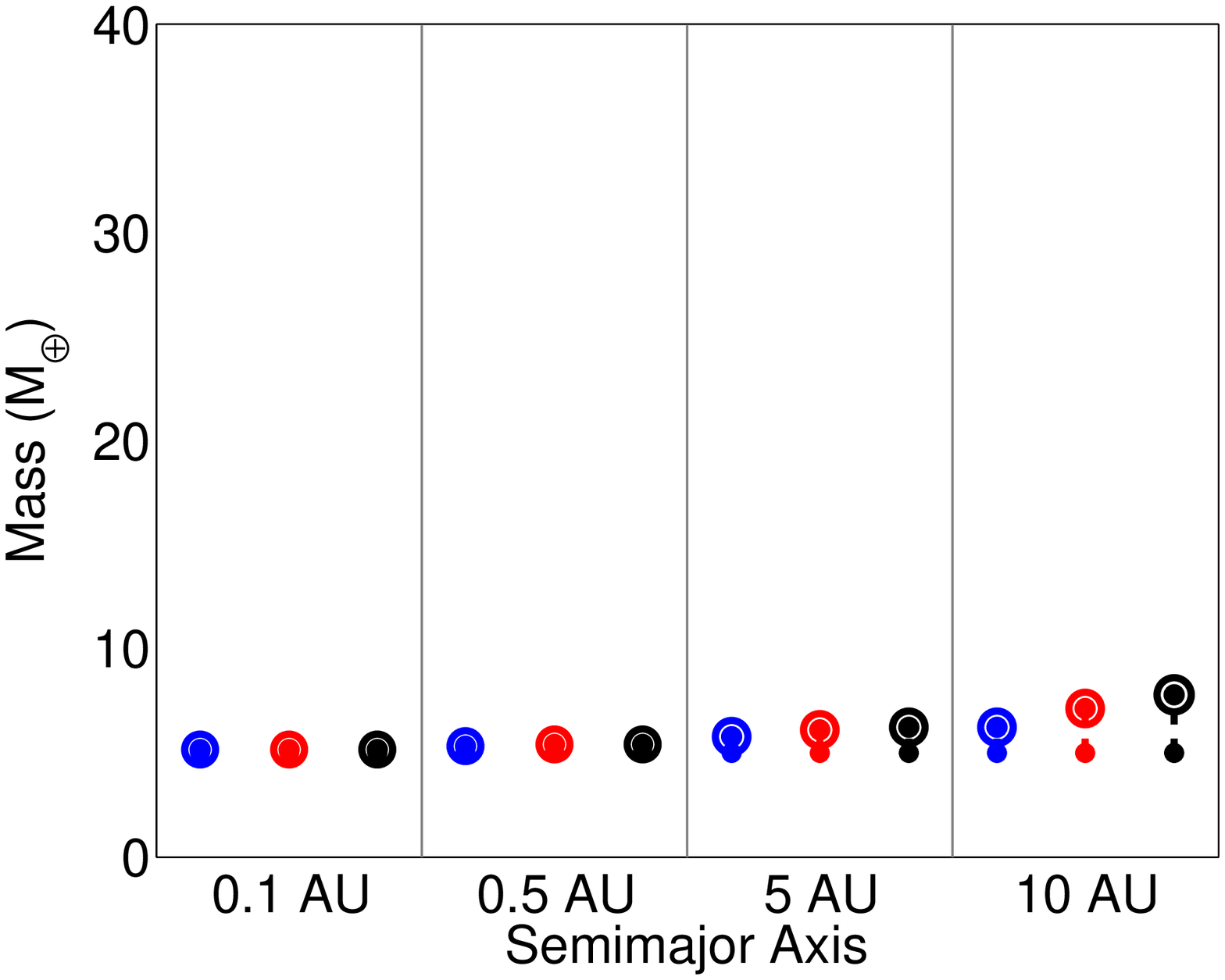}
\includegraphics[width=0.3\textwidth]{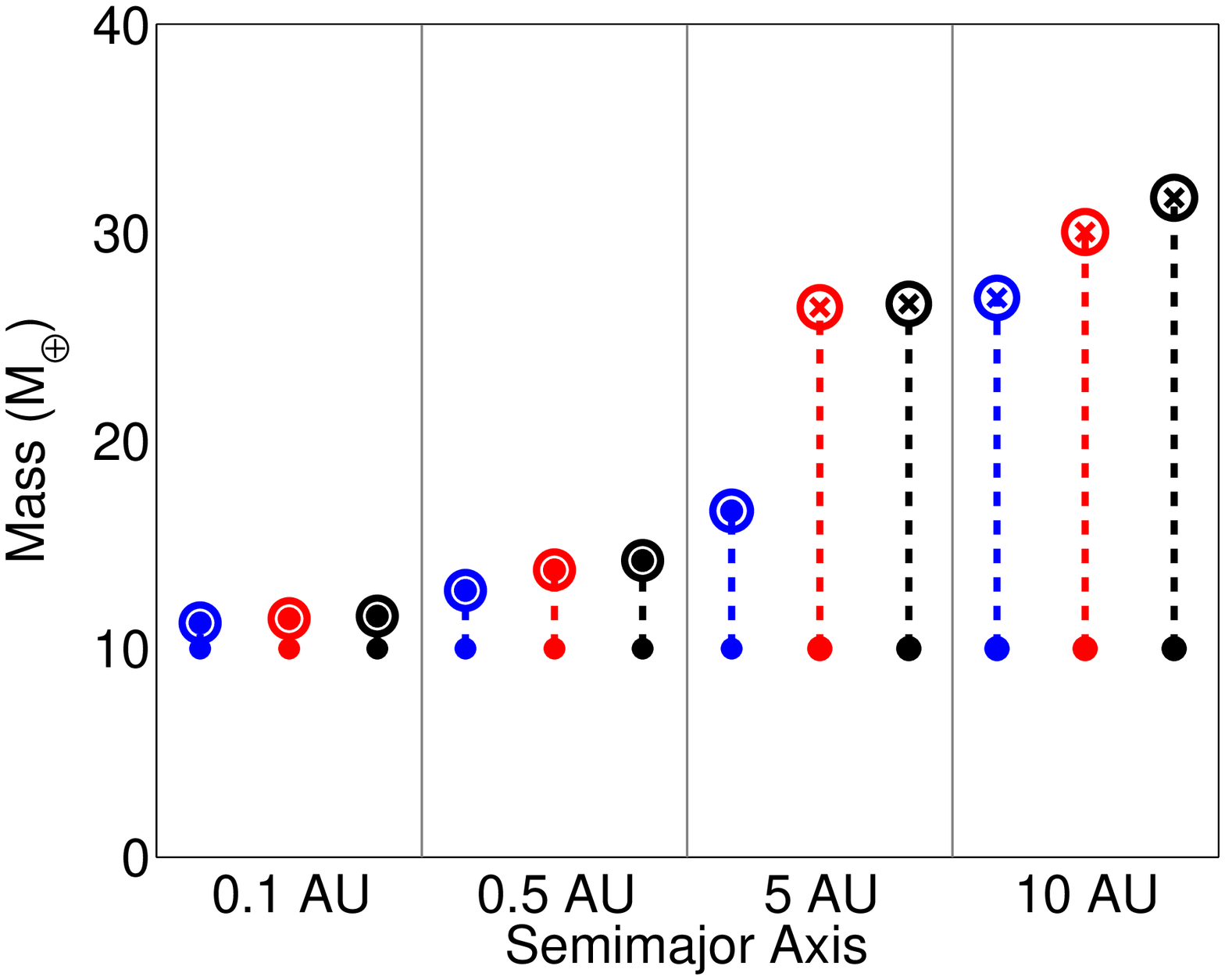}
\includegraphics[width=0.3\textwidth]{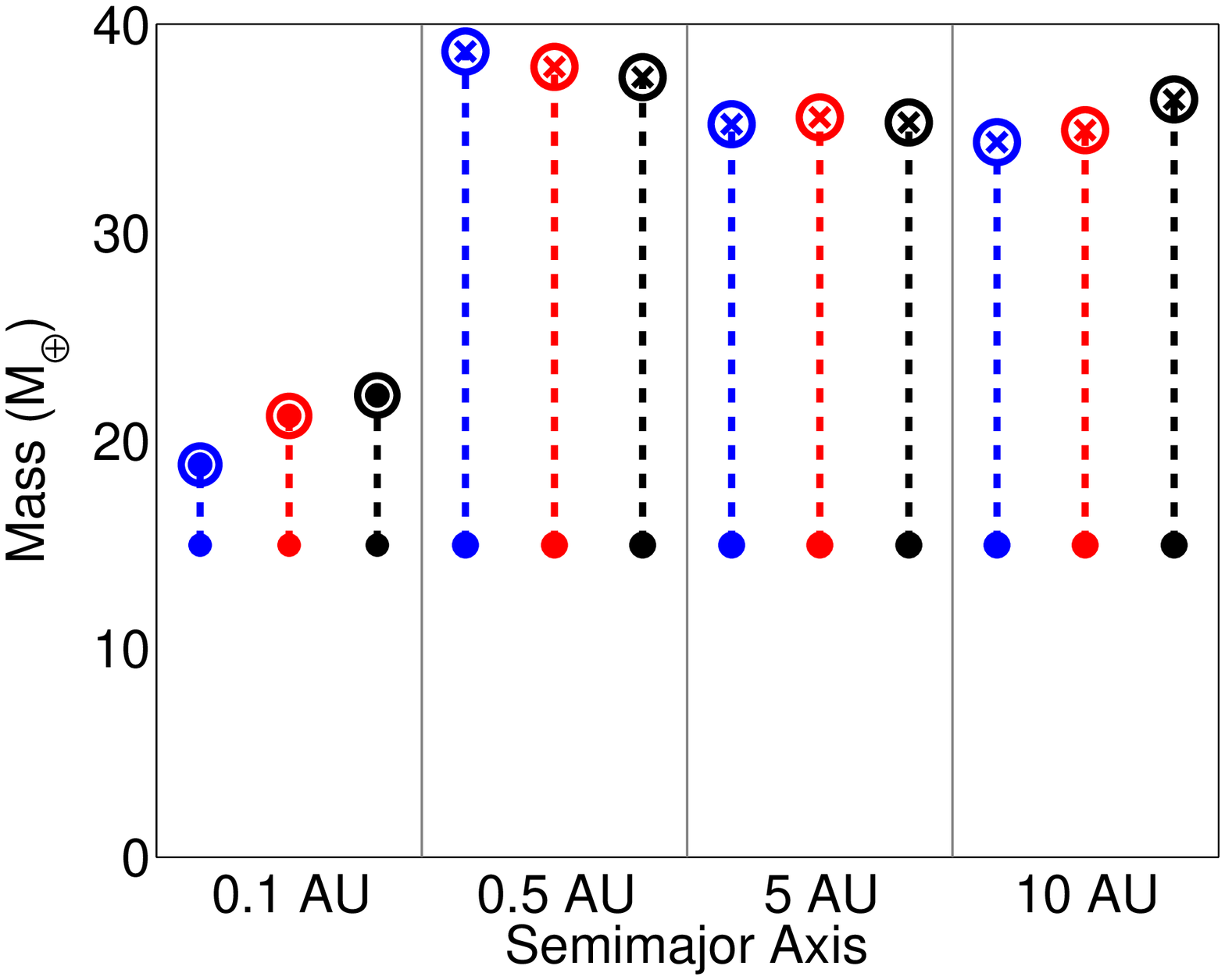}
\caption{Final masses of cores at different locations embedded in evolving discs.
Blue, red and black dashed lines show cores in 1, 3, and 5 $\times \mmsn$ discs respectively.
Different initial core masses are shown in different panels, left-hand panel is for 5 $\me$ cores, centre panel is for 10 $\me$ cores, and right-hand panel is for 15 $\me$ cores.
When a core is denoted by a cross, the envelope has reached a critical state.}
\label{fig:core_evolve}
\end{figure*}

\subsubsection{Accretion at $a_{\rm p} \ge 5 \au$}
The accreting planets located at 0.1 and 0.5 $\au$ discussed above demonstrate how gaseous envelopes settle on to planetary cores of various masses in hot and warm environments close to the central star.
Further out in the disc, however, the temperature is significantly lower, and
this allows for more rapid accretion of massive gaseous envelopes, resulting in the occurrence of runaway gas accretion after the envelopes have reached the critical state.

The bottom left panel of Fig. \ref{fig:sma_evolve} shows that at 5 $\au$, planets with 15 $\me$ cores reach critical states after between 1.2--1.35 Myr, and rapidly accrete the gas from their feeding zones before accreting from the disc at the viscous supply rate, reaching masses in the range $1.5 M_{\rm J} \lesssim M_{\rm p} \lesssim 3 M_{\rm J}$ at the end of the disc lifetime. Planets with 10 $\me$ also undergo significantly more rapid gas accretion, such that those placed in the 3 and 5 $\times \mmsn$ discs undergo runaway gas accretion. A common feature of all of the calculations is the similarity of the quasi-static phase of evolution as a function of the disc model. The disc lifetime is primarily responsible for the variation observed between the disc models. For example, 
all of the 10 $\me$ cores placed in 1, 3 and 5 $\times \mmsn$ discs follow almost nearly identical tracks until the 1 $\times \mmsn$ disc reaches the end of its lifetime, leaving this 10 $\me$ core with a sub-critical gaseous envelope containing $\sim 8 \me$. The cores placed in 3 and 5 $\times \mmsn$ discs, however, reached a critical state after 6.3 and 6.6 Myr respectively, when their masses were 26.1 and 26.6 $\me$, and grew to become gas giants with masses between
0.5--1.5 $M_{\rm J}$. As with the 5 $\me$ cores located at 0.1 and 0.5 $\au$, cores of this mass located at 5 $\au$ failed to accrete significant gas envelopes, with the most massive forming in the 5 $\times \mmsn$ disc with a final envelope mass of 1.3 $\me$.

Finally, the bottom right panel of Fig. \ref{fig:sma_evolve} shows the evolution of cores placed at 10 $\au$ in the disc.
The planets with 15 $\me$ cores all reach the critical state after 0.93--0.95 Myr with total masses $\sim 35 \me$. Due to being located further out in the disc, where the planet's feeding zones are larger, and undergoing runaway gas accretion at earlier times, these giant planets reach masses of between 2--4 $M_{\rm J}$ at the end of the disc lifetime.
The planets with 10 $\me$ cores also reached the critical state once their masses were between 27 and 31 $\me$. These planets all underwent runaway gas accretion.
The planet in the $1 \times \mmsn$ disc entered the runaway state near the end of the disc lifetime, when there was a limited supply of gas available, so this planet did not manage to reach a Jovian mass but instead achieved a final mass of $44 \me$. The planets with 10 $\me$ cores in the more massive discs finished the simulations with masses between 2--3 $M_{\rm J}$.

As with the other locations in the disc, the 5 $\me$ cores again failed to accrete enough gas to undergo runaway gas accretion.

To summarise, cores that accrete gaseous envelopes close to the central star are unable reach a critical state due to the high gas temperatures there, but accretion in the cooler disc regions further from the star is much more efficient. The curves shown in the various panels of Fig. \ref{fig:time_gas_acc} explain this behaviour. Over most of the evolution, the luminosity, energy change per unit mass, and the mass accretion rates are more or less constant. Planet models located close to the star in hot regions of the disc have lower luminosities, larger values (in magnitude) of ($dE /dM_{\rm p}$), and hence smaller accretion rates than planet models located further from the star but with the same core mass. Significant variation of these quantities as a function of time is only really observed at the very beginning of the calculations, and during the time when a planet model approaches the critical state.
During this latter phase, Fig. \ref{fig:time_gas_acc} indicates that the luminosity increases as a result of the increasing and more strongly bound envelope mass having a  higher temperature at the surface of the core which leads to an increased rate of outward heat transport, and ($dE /dM_{\rm p}$) decreases in magnitude sharply, resulting in the dramatic increase in the accretion rates observed in all calculations of this type \citep[e.g.][]{Pollack}.

\begin{figure*}
\centering
\includegraphics[width=0.3\textwidth]{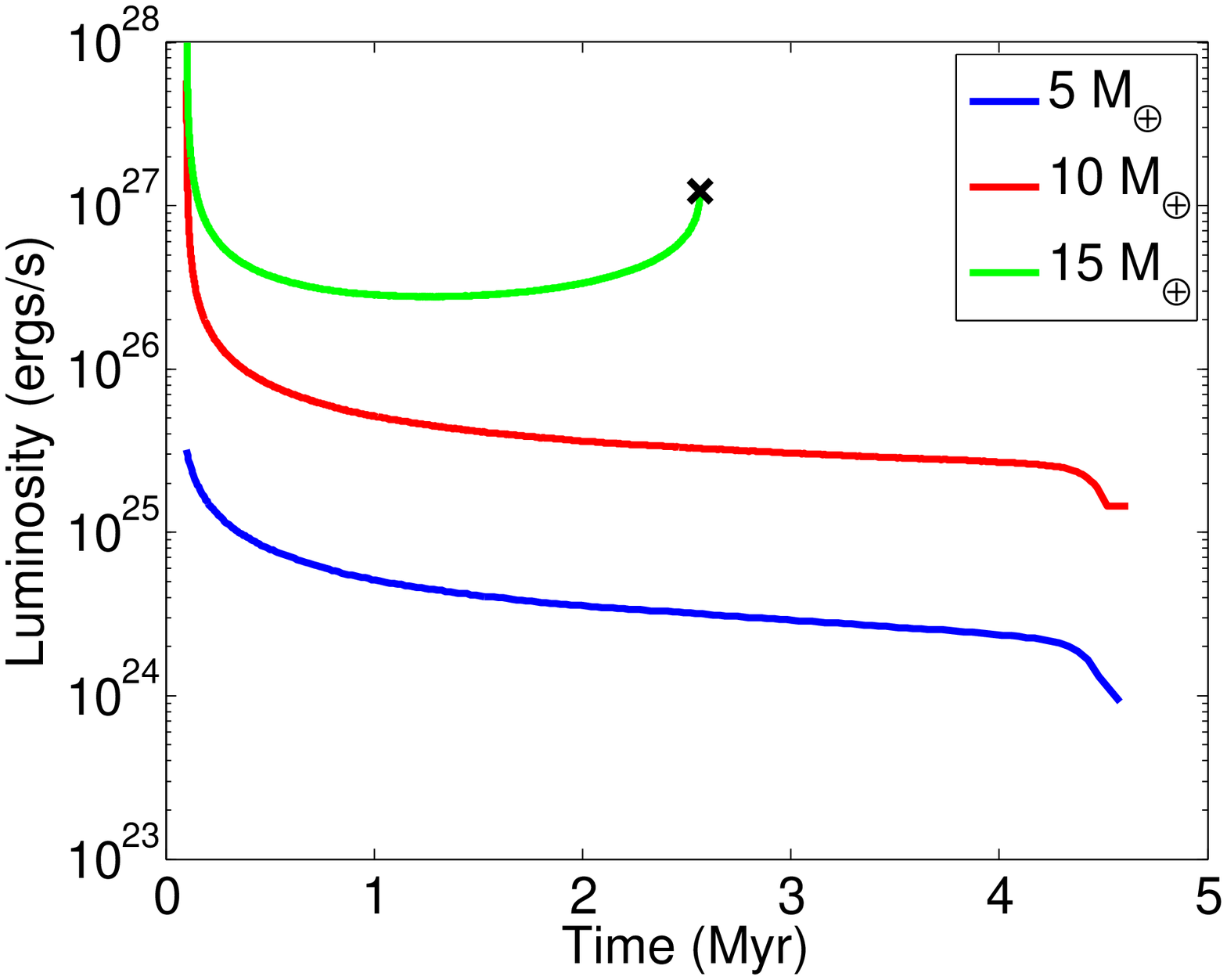}
\includegraphics[width=0.3\textwidth]{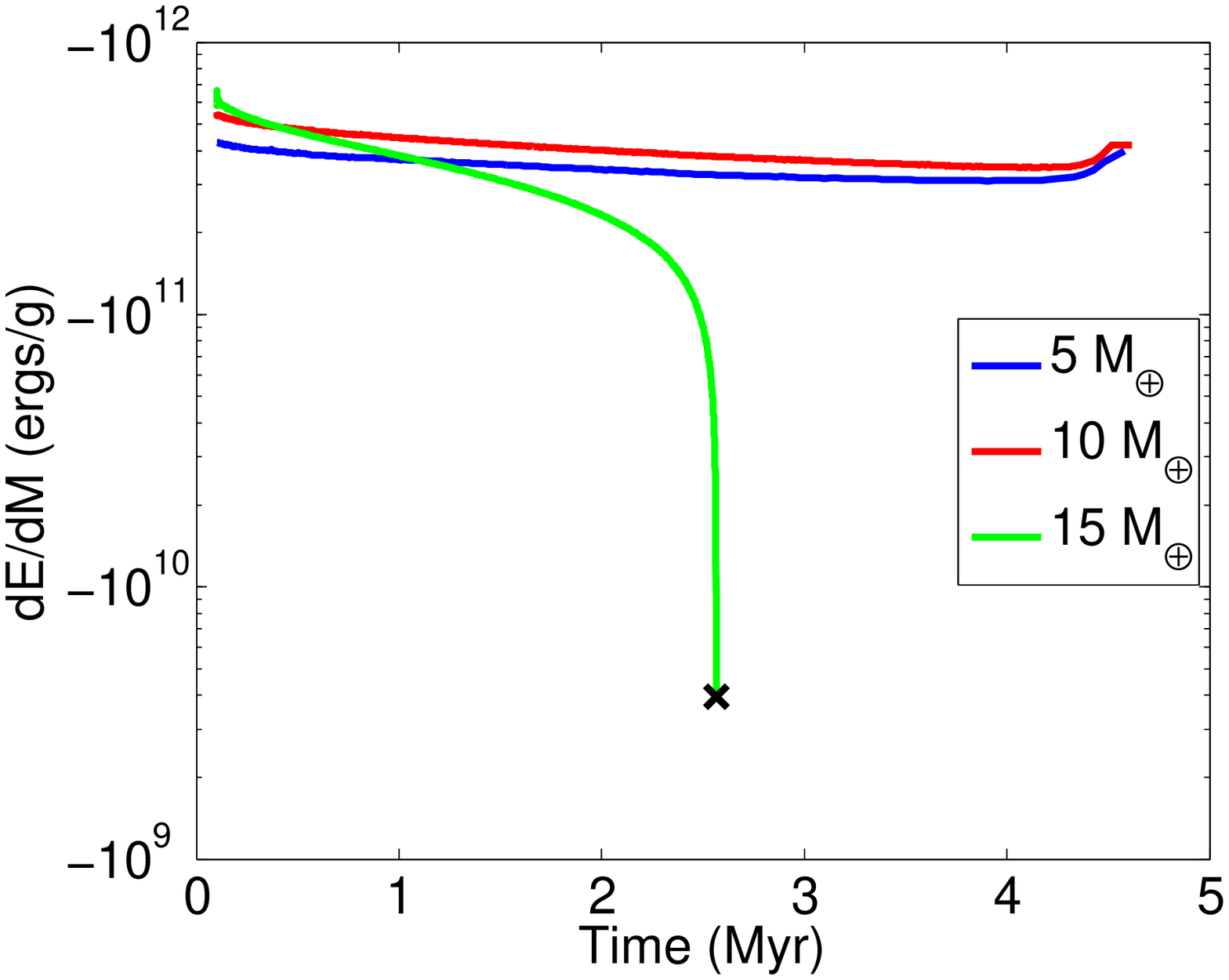}
\includegraphics[width=0.3\textwidth]{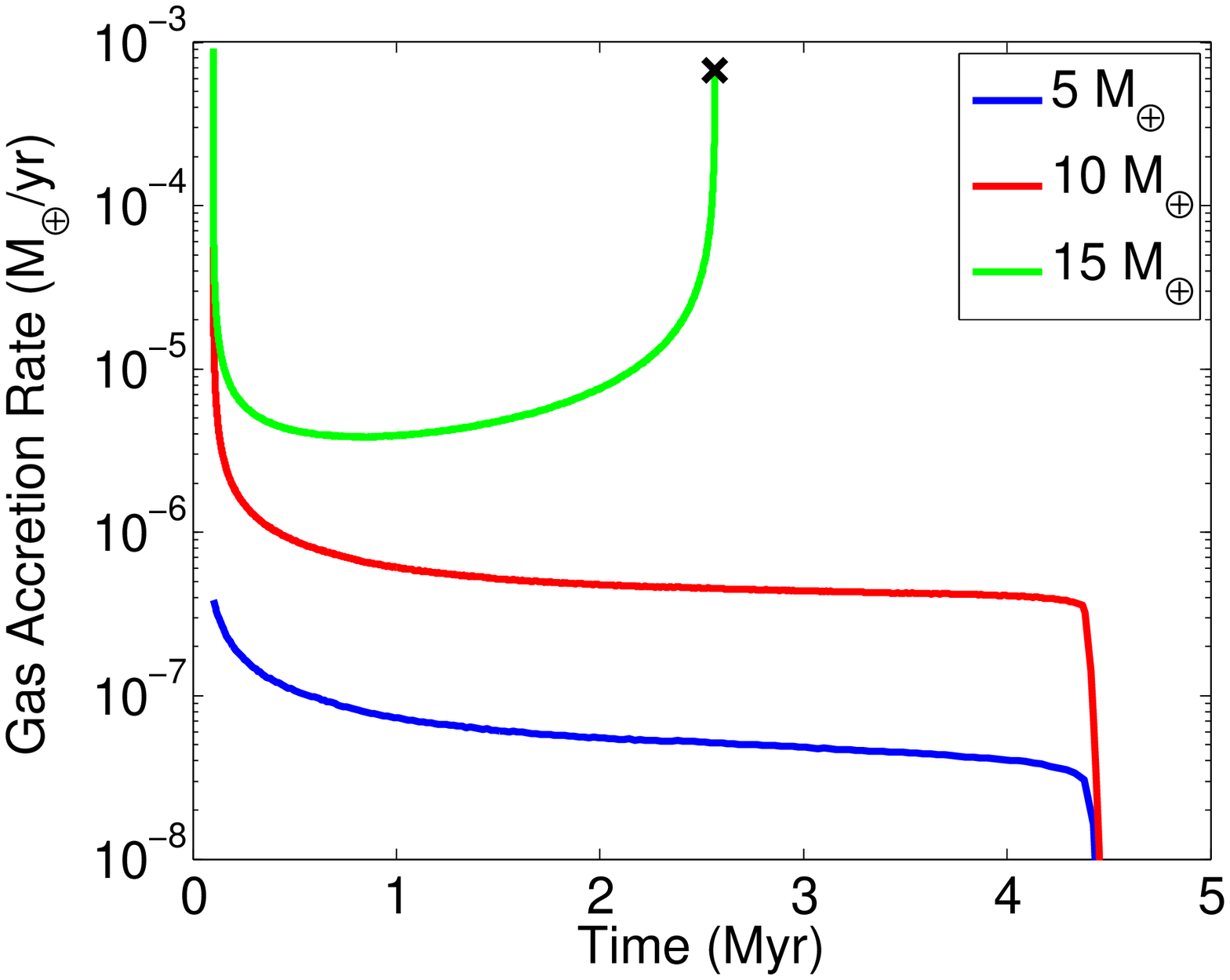}
\caption{Evolution of the planet luminosities (left panel), $d E/d M_{\rm p}$ (middle panel) and mass accretion rates (right panel).
Blue, red and green lines show planets with $5\me$, $10\me$ and $15\me$ cores, respectively, located at $0.5\au$ in the $1 \times \mmsn$ disc.
When a planet's evolution ends with a cross, the envelope has reached a critical state.}
\label{fig:L-dEdM-dMdt-coremass}
\end{figure*}

\subsubsection{Influence of core mass}
\label{sec:core-mass}
The previous sections examined how the accretion and evolution of gaseous envelopes depended on the core's location in the disc.
Figure \ref{fig:core_evolve} shows how the accretion of gaseous envelopes is dependant on the initial core mass, and the general trend of higher core masses resulting in faster gas accretion rates can be seen.
Each panel shows the initial core mass, linked by a vertical dashed line to a final core and envelope mass.
When a core is denoted by a cross, the envelope has reached a critical state, and we plot the mass at this time (not the final mass after runaway gas accretion has occurred).
The blue, red and black colours denote the accretion in 1, 3 and $5 \times \mmsn$ discs respectively, and we show the results of cores accreting gaseous envelopes at 0.1, 0.5, 5 and 10 $\au$.
The left panel shows the evolution of 5 $\me$ cores.
Only at 10 $\au$ in the most massive disc considered here, is such a core able to accrete a sizeable envelope, comprising $\sim 36 \%$ of the total mass.
The middle panel shows the evolution of 10 $\me$ cores, which are able to accrete much more massive gaseous envelopes than their 5 $\me$ core counterparts.
The right panel shows the evolution of 15 $\me$ cores.
The increase in final gas envelope mass with increasing core mass is evident here, with only the planets located close to the star at $0.1 \au$ failing to undergo runaway gas accretion. 
Figure \ref{fig:core_evolve} shows that for planets to accrete significant gaseous envelopes and reach a critical state, they need to have masses of $10 \me$ or higher.
This is in line with results obtained in some different contexts to ours, such as in prescribed nebula models with planetesimal accretion included \citep[e.g.][]{Pollack}.

The trend for increasing gas accretion rate with core mass can be easily understood from inspection of Fig. \ref{fig:L-dEdM-dMdt-coremass}, which shows the evolution of various quantities for planets with different core masses located at $0.5 \au$ in the $1 \times \mmsn$ disc. Planets with more massive cores require larger luminosities to support their envelopes due to the larger binding and thermal energies per unit mass, and collectively the planets have values of $dE/dM_{\rm p}$ which are similar in magnitude during the early and middle phases of gas accretion, leading to larger gas accretion rates and shorter evolution times for planets with more massive cores. 

\section{Long term evolution}
\label{sec:lte}
In order to have results that are comparable with observations on appropriate time-scales, it is necessary to consider the evolution of protoplanets after disc dispersal.
During this phase the envelope contracts in order to release gravitational energy
to balance the energy that is radiated away and the radius shrinks while the planet surface is
heated by radiation from the central star.
We consider models that have not reached a critical state before disc dispersal
and thus only objects that may become Super-Earths or Neptunes.
For initial models we take a sample of protoplanets that were  embedded in their nascent disc during its lifetime
just after its dispersal. Their total mass, core mass  and orbital location at that stage are listed in table \ref{tab:FreeS}.
Models A1, A2 and A3 have $5\me$ cores  and are of increasing envelope mass.
Models B1, B2  and B3 are of increasing envelope mass, each having $10\me$ cores. 

\begin{table}
\centering
\begin{tabular}{lccc}
\hline
Model  &  Total mass  &       Core mass    &        Orbital radius\\
       &  $M_{\rm p}/\me$ &     $ M_{\rm core}/\me $ &   $ a_{\rm p}/au$ \\
\hline
     A1 &   5.216 &                5.00 &           0.2\\
     A2 &   5.398 &                5.00 &           1.0\\
     A3 &   5.776 &                5.00 &           5.0\\
     B1 &   11.821&               10.00 &           0.2\\
     B2 &   13.380&               10.00 &           1.0\\
     B3 &   16.578&               10.00 &           5.0\\
\hline
\end{tabular}
\caption{Initial parameters for  models evolved after disc dispersal}
\label{tab:FreeS}
\end{table}

\begin{figure*}
\centering
\vspace{-8cm}
\hspace{-1cm}
\includegraphics[trim= 100 90 100 170,width=0.70\textwidth]{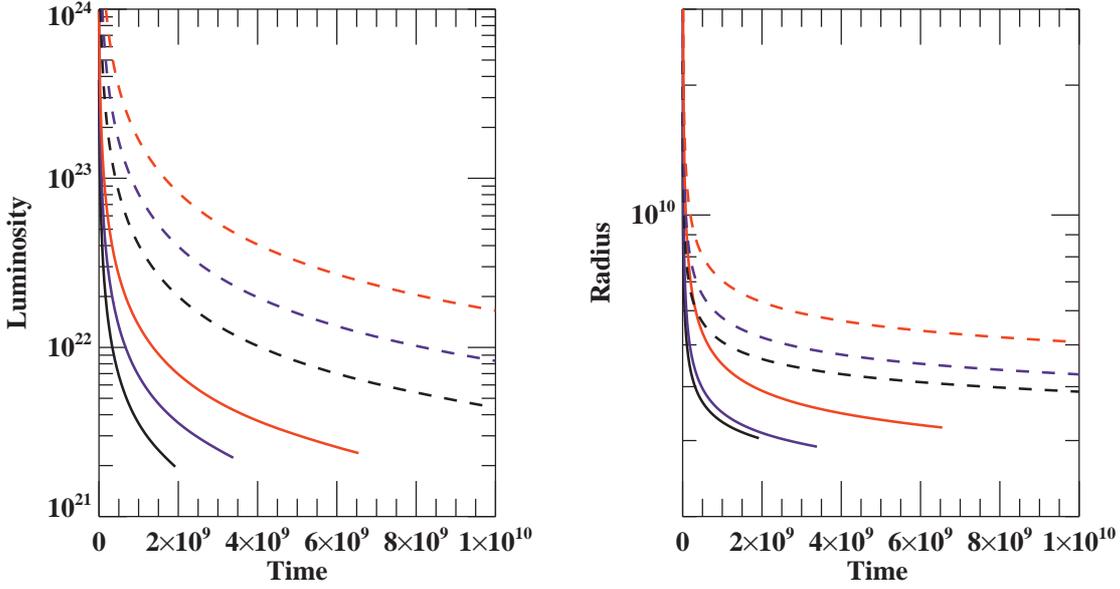}
\caption{The evolution of models with initial parameters given in table \ref{tab:FreeS}.
The internally generated luminosities (left panel-hand) and radii (right-hand panel), in $cgs$ units, are shown as a function of time.
Black, blue and red solid curves respectively correspond to models A1, A2, and A3.
Black, blue and red dashed curves correspond to models  B1,B2 and B3 respectively.}
\label{fig:Freemodels}
\end{figure*}

\begin{figure*}
\centering
\vspace{-8cm}
\hspace{-1cm}
\includegraphics[trim= 100 90 100 170,width=0.7\textwidth]{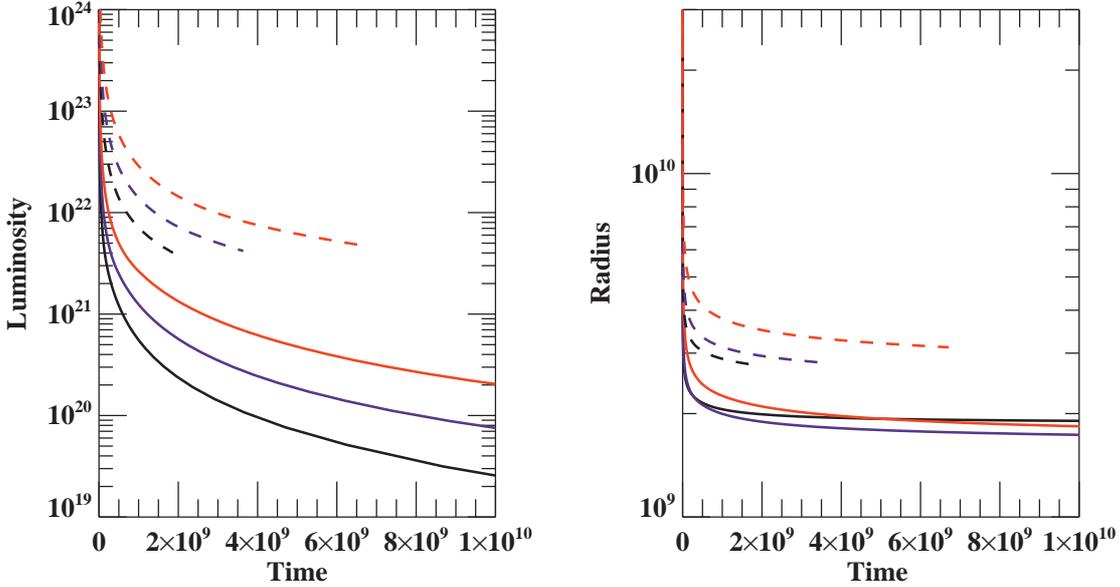}
\caption{As in Fig. \ref{fig:Freemodels} except that  corresponding
models are initiated with envelope masses that are reduced by a factor of five.
\label{fig:FreemodelsS}}
\end{figure*}

The calculations are performed as outlined in Section \ref{sec:freecalc}.
The evolution of the models with initial parameters listed  in table 
\ref{tab:FreeS} is illustrated in Fig. \ref{fig:Freemodels}.
Radii and internally generated luminosities are shown as a function of time.
It will be seen that after an initially fairly rapid contraction
 from the vicinity of the Hill sphere, and reduction in internal luminosity,
the evolution rate slows significantly. The evolution rate decreases with both increasing initial core mass
and increasing envelope mass, the radii of models A1, A2 and  A3 reaching $3\times 10^9$ cm (i.e. $\sim 4.6 \Rearth$) after $\sim 2, 4$ and 8 Gyr, respectively. To investigate the dependence on the initial envelope mass we ran the same models
with their initial envelope masses reduced by a factor of $5.$
The results are illustrated in Fig. \ref{fig:FreemodelsS}.
It will be seen that the evolution times to attain a radius of $3\times 10^9$ cm are reduced to $\sim 0.5$, 3 and 4.5 Gyr, respectively.
This type of evolution indicates that the models may not be fully relaxed
on Gyr time scales, and as a consequence observed systems may exhibit residual inflation.

\begin{table}
\centering
\begin{tabular}{lcc}
\hline
Model  & Final mean density  &  Radius at 2Gyr/ (Final radius)\\
       &  ${\bar \rho}$ &        $ R_{\rm surf}/R_{\rm surf,f}$ \\
\hline
     A1 &  0.674  &                    1.37 \\
     A2 &   1.10 &                     1.63\\
     A3 &   1.13 &                     2.02\\
     B1 &   0.68&                    1.58 \\
     B2 &   0.83&                    1.83\\
     B3 &   0.81&                    1.52\\
\hline
\end{tabular}
\caption{The first column gives the mean density in  the 
final state with zero internal luminosity for each of the models
listed in table (\ref{tab:FreeS}). The second  column
gives the ratio of the radius after 2 Gyr and the 
end state radius.
\label{tab:FreefinalS}}
\end{table}

\subsection{Effect of reducing the opacity}
We have found that the main quantity determining the rate of evolution is the form of the opacity.
This is because this is determined by the rate at which radiation can diffuse out of the optically thick envelope.
To illustrate this we have considered the evolution of models for which the opacity
was reduced by a factor of $100$. The models considered were A1, A2 and B3, though in the latter case we
also reduced the initial envelope mass by a factor of $10$.

\begin{figure*}
\centering
\vspace{-8cm}
\hspace{-1cm}
\includegraphics[trim= 100 90 100 170,width=0.70\textwidth]{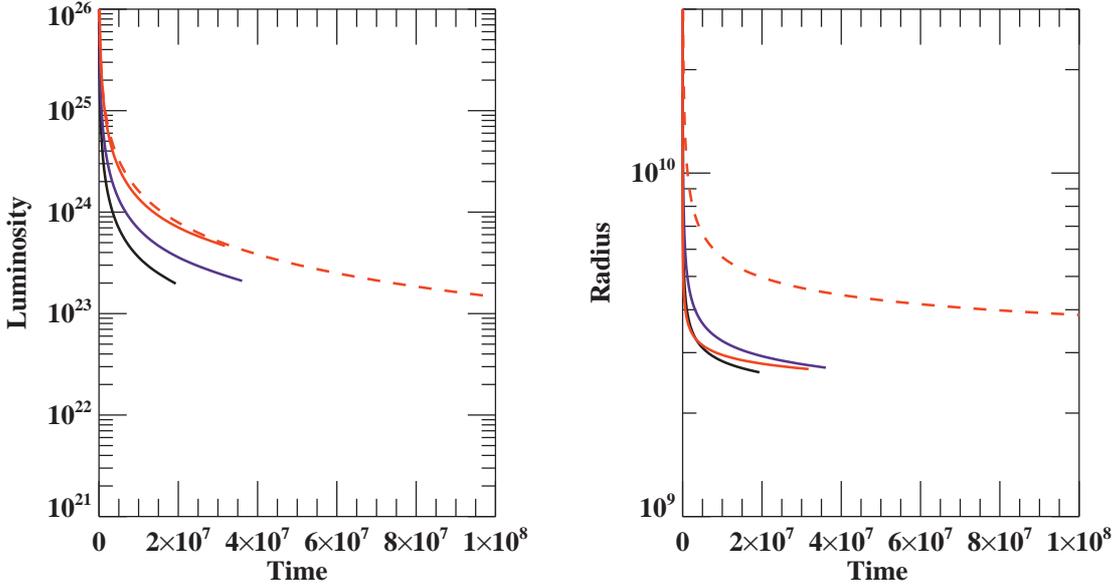}
\caption{The evolution of models with reduced opacity.
Luminosities (left-hand panel) and radii (right-hand panel) are shown.
Black and blue solid  curves correspond to models A1 and A2 with opacity reduced by a factor of 100.
The red dashed curve corresponds to model B3 without opacity reduction
but with initial envelope mass reduced by a factor of 10. The red solid curve corresponds
to the same model but with the opacity reduced by a factor of 100.}
\label{fig:Freemodelsopac}
\end{figure*}

The evolution of these models is illustrated in Fig. \ref{fig:Freemodelsopac}.
Black and blue solid  curves correspond to models A1 and A2 with the opacity reduction.
The red  dashed curve corresponds to model B3 without opacity reduction but with initial envelope mass reduced by a factor of 10.
The red solid curve corresponds to the same model but with the opacity reduction. 
By comparing Figs. \ref{fig:Freemodels} and \ref{fig:Freemodelsopac} we see that the evolution rate for the models with the opacity reduction is sped up by a factor of $100$.
This indicates that the evolution rate $\propto 1/\kappa.$

This may be understood as a consequence of most of the envelope mass being
convective and as a consequence almost isentropic.
At the boundary of the interior convection zone we have $\nabla_{\rm rad}=\nabla_{\rm ad},$
which is equivalent to
\begin{equation}
 \frac{P}{T}\frac{dT}{dP} =\frac {3\kappa P L_{tot}} {64\pi GM_{p}\sigma T^4}=\nabla_{\rm ad}\equiv 
\frac{\Gamma_2-1}{\Gamma_2}.\label{convcond}
\end{equation}
In the exterior radiative region only  the first equality in the above holds.
For an opacity law $\kappa \propto T^{\eta},$ with $\eta < 4,$ this implies that the radiative region tends to a state for which $P \propto T^{4-\eta}.$ Convection will then arise if
\begin{equation}
 \frac{1}{4-\eta}   >
\frac{\Gamma_2-1}{\Gamma_2}.
\end{equation}
In our case $\Gamma_2 \sim 1.4$ so convection ensues if $\eta \gtrsim 1/2.$
This is found to become satisfied with the onset of convection always occurring when $T \sim 1.8\times 10^3K$ (see above).
Given convection is efficient, the model can be assumed to be isentropic to a good approximation.
Equation (\ref{convcond}) and the above discussion will then imply that the structure depends only on the product of the luminosity and the opacity scaling.
As the rate of evolution is $\propto L_{tot}$ (see equation \ref{eq:dmpdtequationfree}) which is determined as a function of the envelope mass in the course of solution, this must then be $\propto 1/\kappa$ as we have found.

Note too that the model radii at corresponding times, taking into account the opacity scaling, are almost independent of this scaling.
Thus we find that model B3 with initial envelope mass reduced by a factor of 10 and the same model with opacity reduced by a factor of 100 differed by < 2\% at corresponding times ( $3\times 10^9$ yr in former case and $3\times 10^7$ yr in the latter).
This indicates insensitivity to the surface boundary condition.

We have seen that the opacity scaling is important for determining the rate of evolution of the isolated planets.
At late stages the envelope becomes cool and dense. For most of the calculations presented above, as has been commonly done \citep[see e.g.][]{Valencia2013}, we have effectively employed an extrapolation of results obtained for much lower densities to obtain opacities.
This is clearly uncertain.
However, in spite of this we can determine the end state of the evolution with the opacity scaling then determining the time taken to attain this.
Consequently, the proximity of any configuration to its end state at a given time is related to the magnitude of the opacity.

\subsection{End states}
As we have seen the above models contract at a progressively lower rate with decreasing luminosity.
The end state is expected to be one with zero internally generated luminosity for which the envelope will be isothermal with a temperature determined by the external radiation field.
We shall assume the envelope is optically thick so that this radiation can be effectively absorbed and provide heating.
As indicated in Section \ref{sec:endstatec} we have determined the end state configurations for the models listed in table \ref{tab:FreeS}.

Results are given in table \ref{tab:FreefinalS} which gives the mean density in  the 
final state  for each of the models.
This is seen to range between 0.67 and 1.13 g cm$^{-3}$.
We also 
give the ratio of the radius after 2 Gyr to that of  the 
end state radius. This number ranges between 1.37 and 2.02,  giving
 an indication of the degree of inflation
that may still exist after that time.
Thus evolving protoplanets may exhibit a range of mean densities and radial
inflation which depend on the initial envelope mass and the form of the opacity,
the latter determining the proximity to the end state. 

\section{Discussion and Conclusions}
\label{sec:conc}
We have presented the results of simulations of gas accretion onto planetary cores that are embedded in the viscously and thermally evolving protoplanetary disc models described in \cite{ColemanNelson16}. The calculations consist of a temporal sequence of hydrostatic planetary envelope models in which the gas accretion rate at each time step is determined by the luminosity of the planet \citep{PapNelson2005}. We have considered in situ gas accretion onto cores with masses of 5, 10 and 15 $\me$ located at orbital radii 0.1, 0.2, 0.5, 1, 2, 5, 10 $\au$ within disc models whose masses are equivalent to 1, 3 and $5 \times \mmsn$, giving rise to a total of 63 models. The aim is to examine how much gas can accrete onto planetary cores as a function of their mass and orbital radius within evolving disc models whose lifetimes span the range that is inferred from observations of young stellar clusters \citep[e.g.][]{Haisch2001}. For simplicity we adopt standard interstellar opacities \citep{Bell94, Bell97} and a standard equation of state \citep{Saumon95} and leave a more complete exploitation of the effects of varying these for future work. For a subset of the planet models that have not undergone runaway gas accretion at the time of protoplanetary disc dispersal, we evolve the planets forward in time for Gyr time scales to follow their long term cooling and contraction.
Our main results can be summarised as follows. \\
(i) For a fixed core mass and disc model, gas envelope accretion occurs more rapidly further from the central star where the gas temperature is lower. This is in agreement with recent studies of gas accretion onto cores located at large distances from the star which also reached the same conclusion \citep{Piso2014, Piso2015}. We also find that during the slow gas accretion phase, which occurs prior to runaway gas accretion, the mass growth histories of planets at fixed orbital radii only vary very slightly between the different disc models, with higher mass discs generally favouring slightly more rapid gas accretion.
Only once the planets undergo runaway gas accretion and begin to accrete gas from the disc at the viscous supply rate, do the mass evolution tracks differ substantially.
These differences arise due to the remaining disc mass and lifetime, where giant planets undergoing runaway gas accretion in more massive discs have more time to increase their masses than those that start to undergo runaway gas accretion at roughly the same time in lower mass discs.\\
(ii) None of the planets that we considered were able to undergo runaway gas accretion at an orbital radius of $0.1 \au$, due to the long accretion times that we obtained in this hot region of the disc where the temperature remains close to 1000 K throughout most of the evolution. At this temperature the opacity of the outer radiative zone of the planetary envelope remains dominated by dust grains, and so the Kelvin-Helmholtz contraction of the envelope takes longer than the disc model lifetimes of between 4.6 Myr and 9.4 Myr. There have been other recent studies of in situ gas accretion that have resulted in planets undergoing runaway gas accretion close to the star that at first sight appear to contradict our results. \cite{Batygin2016} presented models in which $15 \me$ cores located at $0.05 \au$ underwent runaway gas accretion within $\sim 1$ Myr, but these models were computed under static nebula conditions quite different to ours. Their disc temperature remained at 1500 K, above the sublimation temperature of grains such that they did not contribute to the opacity, and it is well known from previous studies that opacity reductions increase gas accretion rates onto planetary cores \citep[e.g.][]{Hubickyj2005,PapNelson2005,Movs,Mordasini2014}. A significantly larger local gas disc density was also assumed. We note, however, that \cite{Batygin2016} also considered gas accretion onto 4 and $10 \me$ cores, and in these cases found that only moderate gas accretion occurred, in accord with our results. In situ gas accretion at $0.1 \au$ has also been considered by \cite{Lee2014}, resulting in more rapid gas accretion onto $10 \me$ cores than we observe, again under nebula conditions very different to ours. In their study a static version of the so-called minimum mass extrasolar nebula \citep{ChiangLaughlin2013} was adopted, which has a much higher density and pressure at $0.1 \au$ than our viscously and thermally evolving models, so it would appear that the differences in outcomes and conclusions in their study arise because of the different model assumptions.
In particular we remark that the large optical depths of cores in such very massive discs tends to increase the temperature at the disc-protoplanet interface, which in turn acts to drive cores to criticality more easily than those indicated above.
\\
(iii) As we move further from the star, the planets with 10 and $15 \me$ cores start to undergo runaway gas accretion and become gas giant planets. The planets with $15 \me$ cores all undergo runaway gas accretion for orbital radii $\ge 0.5 \au$, and once the orbital radius reaches $5 \au$ all but one of the $10 \me$ cores experience runaway gas accretion. Taking these results at face value, without considering how reductions of the opacity or other factors could modify the results, we conclude that in situ formation of gas giant planets at $0.1 \au$ without migration probably requires the presence of a core that exceeds $15 \me$, but formation at orbital radii $\sim 0.5 \au$ can occur for core masses $\sim 15 \me$. Viewed within the context of our models, the existence of hot Jupiters orbiting with short periods can either be explained as a consequence of formation further from the star via gas accretion and further solid accretion onto cores with masses $\sim 10$--$15\me$, followed by migration to their current locations, or by in situ gas accretion onto cores more massive than $15 \me$. Both of these scenarios are consistent with the typical core/heavy element masses of $\ge 20 \me$ inferred to exist within hot Jupiters by the numerous studies that have examined this point \citep{Guillot2006, Burrows2007, MillerFortney2011, Thorngren2016}. \\
(iv) None of the $5 \me$ cores in our simulations accreted enough gas to undergo runaway gas accretion at any location in the disc. An obvious conclusion to draw from this is that the systems of short-period Super-Earths and Neptunes that have been observed in abundance \citep[e.g.][]{Fressin2013} probably accreted moderate gas envelopes without undergoing runaway gas accretion, either in situ or further from the star followed by inward migration, because their solid cores have masses $\le 10 \me$, or because the cores accreted to larger values late within the disc lifetime. If this conclusion is correct, then there is likely to be an as yet undiscovered population of Super-Earths/mini-Neptunes orbiting at large radii with moderate gas envelopes sitting on cores with masses $\lesssim 5 \me$. \\
(v) The mean densities of those planets that were evolved for Gyr time scales after the dispersal of the gas disc reach values of $\bar \rho \sim 0.1$--0.3 g cm$^{-3}$ after contracting for 2 Gyr. The final end states to which these planets evolve on longer time scales have mean densities in the range $\bar \rho \sim 0.7$--1 g cm$^{-3}$. This range of values encompasses the range of densities inferred for the Solar System gas and ice giants, and the values inferred for the super-Earths and Neptunes among the exoplanet population \citep{Marcy2014, Weiss2016}.

\subsection{Future Work and Direction}
Whilst this paper addresses the issue of accreting gaseous envelopes on to pre-existing cores in situ, there are many areas of parameter space and model improvements that we intend to explore in future work. These include:

\noindent (i) Allowing the planets to migrate whilst accreting gaseous envelopes.
Planets of the masses described here can have short migration times relative to disc lifetimes, which could result in planets not having time to undergo runaway gas accretion before they migrate into the star, unless they form at very large radii and/or experience some kind of migration stopping mechanism.
Another effect of considering migration would be the ever changing disc conditions as the planet migrates from a cold region to a hot one (or vice versa). Rapidly changing conditions could have significant consequences for the evolution of the atmosphere and the subsequent accretion rates.

\noindent (ii) Including a stochastic accretion rate for the core's mass through the introduction of planetesimals/boulders/pebbles into the models.
With the accretion of macroscopic solids, there will also be an accretion luminosity which will contribute to the luminosity of the planet.
This contribution to the luminosity will affect the gas accretion rate, and subsequently the structure of the envelope.
But note that, as indicated in \ref{sec:atmos_accretion}, this will lead to lower envelope masses than those found here because we assumed planetesimal accretion to be inoperative.
Some of the very large heavy element abundances inferred for gas giant exoplanets such as CoRoT 10-b, HD 80606-b and HD 149026-b \citep{Guillot2006, Burrows2007,MillerFortney2011,Thorngren2016} suggest that heavy elements are likely to be present in both the cores and the envelopes of some planets, such that this model extension is required to account for at least some fraction of the giant exoplanet population.

\noindent (iii) Combine the models presented here with ab-initio models of planet formation similar to those presented in \citet{ColemanNelson14,ColemanNelson16,ColemanNelson16b} to examine the types of planets and planetary systems that emerge with more realistic gas accretion models.

\noindent (iv) Improve the modelling of the long term evolution of planets after dispersal of the protoplanetary disc through consideration of opacity changes and the heat capacity of the core that may provide a source of long term heating for the planets as they cool and contract.

\section*{Acknowledgements}
The simulations presented in this paper utilised Queen Mary's MidPlus computational facilities, supported by QMUL Research-IT and funded by EPSRC grant EP/K000128/1. This research was supported in part by the National Science Foundation under Grant No. NSF PHY-1125915. We acknowledge the referee, Kaitlin Kratter, whose comments helped to improve this paper.

\bibliographystyle{mnras}
\bibliography{references}{}

\end{document}